\documentclass[journal]{IEEEtran}
\usepackage{etoolbox}
\usepackage[draft,protrusion=false]{microtype}
\usepackage{multirow}
\usepackage{graphicx}
\usepackage{subcaption}
\usepackage{booktabs}
\usepackage{cite}
\usepackage{soul}
\usepackage{kotex}
\usepackage{xspace}
\usepackage{mathtools}
\usepackage{algorithm, algorithmic}
\usepackage{balance}
\usepackage{url}
\usepackage{amsfonts}
\usepackage{nicefrac}
\usepackage{xcolor}
\usepackage{wrapfig}
\usepackage{enumitem}
\usepackage{float}
\usepackage{import}
\usepackage{physics}
\usepackage{bm}
\usepackage{bbm}
\usepackage{amsmath}
\usepackage{amsthm}
\usepackage{amssymb}
\usepackage{qcircuit}
\usepackage{array}
\usepackage[algo2e, linesnumbered, ruled, vlined]{algorithm2e}
\usepackage{tabularx}
\usepackage{hyperref}
\usepackage{mdframed}
\usepackage[normalem]{ulem}
\usepackage[compatibility=false]{caption}

\newcommand{\BfPara}[1]{{\noindent\bf#1.}\xspace}

\newcommand{\T}{\mathcal{T}}

\newlength{\lblSets}
\newlength{\lblPars}
\newlength{\lblVars}
\begin{document}
\title{Hierarchical Multi-Agent Reinforcement Learning for Carbon-Aware AI Data Centers in Power Distribution Systems}

\author{
Hyunsoo Lee, Panggah Prabawa,~\IEEEmembership{Graduate Student Member,~IEEE}, Dae-Hyun Choi,~\IEEEmembership{Member,~IEEE}, and 
\\ Joongheon Kim,~\IEEEmembership{Senior Member,~IEEE}
\thanks{H. Lee is with LG Electronics, 17709, Pyeongtaek, Gyeonggi-do, Republic of Korea (e-mail: hyunsoo11.lee@lge.com).}
\thanks{J. Kim is with the Department of Electrical and Computer Engineering, Korea University, Seoul 02841, Republic of Korea (e-mail: joongheon@korea.ac.kr).}
\thanks{P. Prabawa and D.-H. Choi are with the School of Electrical and Electronics Engineering, Chung-Ang University, Seoul 06974, Republic of Korea (e-mails: \{panggah22, dhchoi\}@cau.ac.kr).}
\thanks{D.-H. Choi and J. Kim are the corresponding authors of this paper.}
}

\maketitle

\begin{abstract}
Eco-friendly energy management for artificial intelligence data centers (AIDCs) is crucial because of the significant increase in energy consumption-induced carbon emissions from AIDCs resulting from the rapid expansion of AI applications. This paper proposes a hierarchical carbon-aware multi-agent reinforcement learning (CA-MARL) framework for robust and efficient operations of AIDCs under uncertainties while ensuring low-carbon operation of power distribution systems. The framework comprises a workload manager (WM) agent and multiple local AIDC agents trained using a multi-agent transformer method, corresponding to a global AIDC aggregator and a local AIDC operator, respectively. Leveraging AIDC operation data along with nodal carbon intensity (NCI) calculated from the carbon emission flow-integrated distribution system operator problem, the WM agent spatially allocates AI training and inference jobs among all AIDCs. Based on the jobs allocated from the WM agent and NCI information, each AIDC agent  schedules economical and eco-friendly operations of the AIDC by performing the following tasks: i) temporal shifting of training jobs, ii) spatial allocation of training graphics processing unit (GPU) blocks and inference GPUs within the AIDC, and iii) control of the supply air temperature of the cooling system. The effectiveness of the proposed framework was assessed using an IEEE 33-node power distribution system.
\end{abstract}


\IEEEpeerreviewmaketitle
\section{Introduction} \label{sec:introduction}
\IEEEPARstart{T}{he} rapid proliferation of large language models (LLMs) and generative artificial intelligence (AI) has significantly increased the energy demand of AI data centers (AIDCs). For AIDCs, training a single large-scale model can require several MWh of electricity~\cite{strubell2019energy}. Moreover, heterogeneous global-scale inference workloads maintain a persistent baseline energy consumption~\cite{luccioni2023power}. By 2024, AIDCs accounted for approximately 1.5\% of global electricity use, with expectations of doubling this percentage by 2030~\cite{iea2024datacenter}. This surging energy consumption of AIDCs generates a substantial carbon footprint and may lead to operational instability in the power distribution grids. Therefore, efficient and eco-friendly energy management of power distribution grids and AIDCs, utilizing distribution system operators (DSOs) and AIDC operators, respectively, is becoming increasingly critical.

Compared with conventional Internet data centers (IDCs), AIDCs are characterized by GPU-centric computing environments, higher power density, and AI training workloads that provide greater temporal and spatial scheduling flexibility. Unlike the IDCs, which are independent of AI applications, AIDCs manage two types of AI workloads with different characteristics: training and inference workloads \cite{Chen2025ElectricityDemand}. The former refers to delay-tolerant workloads (e.g., deep learning training) that can tolerate relaxed service delays. For these workloads, temporal shifting is defined as a scheduling mechanism that defers the execution of jobs from high-cost periods to more favorable future time slots to optimize energy consumption. The latter indicates delay-sensitive workloads (e.g., searching) that require an immediate response. For such workloads, spatial allocation is defined as the process of distributing computational tasks across physically dispersed AIDCs to leverage regional variations in carbon intensity and power availability. This spatial and temporal flexibility of AIDCs can be leveraged to maintain economic operations. 
In this study, spatial flexibility refers to the capability of distributing AI training and inference workloads across geographically distributed AIDCs to exploit location-dependent differences in grid operating conditions. While spatial flexibility is typically associated with inference workloads, it can also be leveraged for training workloads in geographically distributed AIDCs, enabling more flexible and carbon-aware operation. However, scheduling only the operation of AIDCs without considering distribution grid operations can have a detrimental impact on the stable and low-carbon operations of distribution grids. Furthermore, the random arrival of various workloads causes uncertain AIDC scheduling in dynamically changing environments. This uncertainty can result in undesirable coordination between DSO and AIDC operators. This leads to unstable distribution grid operations and high carbon emissions. To address these issues, we propose a carbon-aware multi-agent reinforcement learning (CA-MARL) framework in which DSO and AIDC operators interact to achieve stable and low-carbon distribution grid operations while guaranteeing efficient AIDC operation.

Numerous studies have investigated the interactions between IDCs and distribution systems by focusing on cost minimization and reliability enhancement.
An Internet service company--DSO collaborative optimization model minimized the operational costs of the distribution grid and IDCs via the spatial scheduling capability of online workloads of IDCs \cite{dcpower_tsg1}. Furthermore, an incentive-compatible demand response strategy for spatially coupled IDCs was developed, employing an optimal power flow model and a benefit redistribution mechanism to align social welfare with IDC profits in electricity markets \cite{idc_elec}. In addition, the spatio-temporal flexibility of geo-distributed IDCs in virtual power plants (VPPs) was leveraged for IDCs to track the target load curves more precisely, thereby improving demand response performance and maximizing the profit of each VPP \cite{dcpower_tsg2}. A mixed-integer second-order conic programming-based service restoration method using IDCs was proposed in which the DSO minimizes the total losses of dropped workloads and unrestored electric loads by shifting workloads spatially and temporally under extreme fault conditions \cite{spatiotemporal_tsg4}. A privacy-preserving distributed optimization method was proposed for energy management of hierarchical data centers (edge, fog, and cloud data centers) \cite{optimization_tsg}. In this method, the DSO minimizes both data center operation cost and power distribution losses while protecting their private data \cite{optimization_tsg}. More recently, a reinforcement learning (RL)-based power capping scheme for IDCs was developed in which an agent minimizes the energy consumption costs of IDCs and service level agreement violations in an uncertain and dynamic environment without requiring workload-level information \cite{dcpower_tsg3}. Concurrently, under the dynamically fluctuating carbon intensity of distribution grids, various studies on carbon-aware scheduling of IDCs via temporal shifting of delay-tolerant workloads were performed to reduce carbon emissions. These include the impact analysis of temporal workload shifting on carbon emission reduction in Germany, Great Britain, France, and California~\cite{carbonworkload2}, the construction of virtual capacity curves that automatically manage workload scheduling based on carbon intensity forecasting~\cite{radovanovic2024carbon}, and the development of a holistic framework for simultaneously optimizing the locations of renewable energy and battery storage, workload scheduling, and infrastructure investments to achieve carbon-free operation~\cite{acun2023carbon}.
However, these studies primarily focused on conventional IDCs with homogeneous workloads and do not capture the unique characteristics of AI workloads.

Compared to the extensive body of research on IDCs, only a limited number of studies have begun to investigate the scheduling of training and inference workload for AIDCs. For example, a MARL-based approach was proposed to optimize AI-Generated Content-based training workload scheduling across geo-distributed data centers by jointly considering graphics processing unit (GPU) utilization, energy cost, and carbon emissions
\cite{Zhang2023AIDC}. In addition, AI inference workloads were addressed using a game-theoretic deep RL method, which enables efficient and adaptive workload distribution while reducing operational costs and carbon emissions under dynamically varying system conditions \cite{Hogade2025AIDC}. More recently, a realistic benchmarking simulation framework for AI workload management was developed, which captures the complex interactions among varying electricity prices, carbon intensity, weather conditions, and network dynamics in geo-distributed data centers \cite{Antonio2025AIDC}. These recent AIDC studies indicate that scheduling of training and inference workloads has begun to receive increasing attention. However, the integration of AIDC workload scheduling with grid-aware carbon signals and scalable coordinated control remains limited.

The aforementioned challenges, including spatio-temporal workload scheduling, workload uncertainty, and multi-agent coordination between the DSO and AIDC operators, result in a large-scale and complex decision-making problem. To address this complexity, hierarchical reinforcement learning (HRL) has been widely employed  to address the high-dimensional state and action spaces in power system applications. Previous studies have applied HRL to various grid applications by decomposing complex decision-making problems into hierarchical structures. For example, HRL was applied to power network topology control by decomposing the action space into multiple levels to improve scalability in large combinatorial settings \cite{Manczak2023HRL}. Similarly, a hierarchical deep RL framework was proposed for emergency control, where critical actions are first identified and then sequentially executed to enhance stability and convergence performance \cite{Chen2024HRL}. From an energy management perspective, HRL was also utilized for microgrid economic dispatch, where domain knowledge is incorporated to improve learning efficiency and ensure feasibility under operational constraints \cite{Lu2023HRL}. More recently, a MARL approach was developed for coordinated control of building energy systems, such as heating, ventilating, and air-conditioning systems and distributed energy resources, to reduce operational costs and peak demand \cite{Pei2025HRL}.

However, the aforementioned studies present several challenges. First, previous studies on coordinated operation between data centers and power systems mainly addressed the energy management of traditional IDCs, without modeling the complex operation of AIDCs with heterogeneous  training and inference workloads. Second, the carbon-aware scheduling framework of IDCs considered a location-based carbon intensity signal of the grid, which quantifies only the impact of power generation on carbon emissions, without reflecting the physics of power flows and congestion. Recently, a carbon emission flow (CEF) method based on the notion of nodal carbon intensity (NCI) was developed to assess carbon emissions associated with both power generation and consumption in dynamically changing grid conditions \cite{Kang2015TSG}. The NCI-integrated CEF has been widely adopted to reduce carbon emissions in various power system applications, including the planning of an integrated electricity-hydrogen-gas system \cite{Wei2022CEF},  peer-to-peer
energy trading \cite{Liu2025CEF}, and the scheduling of electric vehicle aggregators \cite{Shi2024CEF}. NCI specifies the responsibility of AIDCs as consumers of carbon emissions. Although the CEF model has been proposed to quantify carbon emissions associated with electricity consumption in power systems, its integration with scalable workload  allocation and scheduling frameworks for geographically distributed AIDCs has not yet been investigated. Third, no previous studies addressed the uncertainties in the training and inference workload arrivals for AIDCs within a coordinated framework between the distribution grid and AIDCs. Fourth, although HRL has been widely employed to resolve high-dimensional decision-making problems in power system applications, existing HRL-based approaches are limited to single-domain operational problems. They do not consider cross-domain coordination between distribution systems and geographically distributed AIDCs. In such settings, spatio-temporal workload scheduling, workload uncertainty, multi-agent interactions, and CEF-based NCI signals significantly increase the complexity of decision-making. Therefore, scalable hierarchical frameworks for economical and eco-friendly DSO--AIDC operator coordination remain unexplored.

\begin{figure}[t!]
    \centering
    \includegraphics[width=0.99\linewidth]{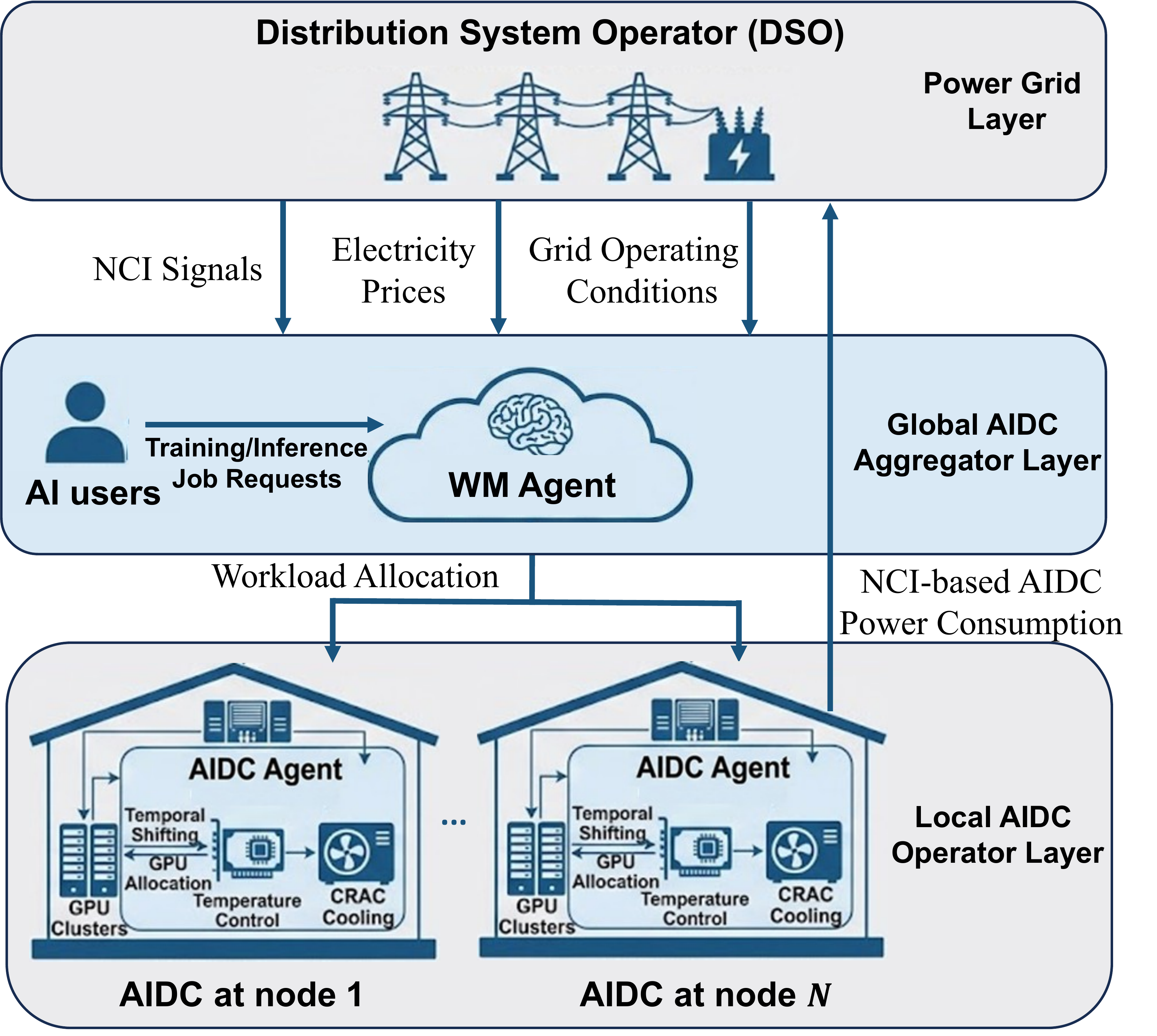}
    \caption{A reference hierarchical system integrated with a DSO, a WM agent, and AIDC agents.}
    \label{fig:system}
\end{figure}

To address these challenges, we propose a hierarchical CA-MARL framework for robust and efficient DSO--AIDC operator coordination in uncertain AIDC operational environments.
As shown in Fig.~\ref{fig:system}, the proposed framework consists of three entities: i) a DSO, ii) a global AIDC aggregator, and iii) local AIDC operators.
In the proposed framework, the global AIDC aggregator is implemented as the workload manager (WM) agent, while the local AIDC operators are represented by AIDC agents, and the DSO is modeled as the environment with which the WM and AIDC agents interact.
Motivated by distinct workload scheduling responsibilities and the need to prevent combinatorial action-space explosion, we propose a hierarchical structure. Specifically, the WM agent for the global AIDC aggregator and AIDC agents for the local AIDC operators function hierarchically. These agents continuously interact with the environment associated with the DSO. There is no direct information exchange among AIDC agents, and their coordination is achieved through the WM agent within the hierarchical structure. Based on the varying grid and AIDC operating conditions, including NCI signals, the WM agent spatially allocates AI training/inference workloads across geographically distributed AIDCs. 
Each AIDC agent calculates its optimal power-consumption schedule using the workloads allocated by the WM agent. To achieve this, the agent performs temporal shifting of training jobs, spatial allocation of graphics processing units (GPUs) for workloads, and control of the supply air temperature of the computer room air conditioning (CRAC)-based cooling system.
Finally, the power consumption schedules calculated by all AIDC agents are sent back to the DSO environment.
In the proposed framework, the DSO is responsible for maintaining the operational constraints of distribution system, while the WM and AIDC agents perform workload allocation and scheduling based on NCI signals. 

Among various MARL methods, a multi-agent transformer (MAT) method is adopted in this study because the proposed carbon-aware AIDC scheduling problem involves hierarchically coordinated yet strongly coupled multi-agent decisions under a rapidly growing joint action space.

The main contributions are summarized as follows:
\begin{enumerate}
    \item We present a closed-loop CA-MARL model for DSO--AIDC operator coordination in which the DSO and AIDC operators cooperate to realize economical and low-carbon distribution grid and AIDC operations by iteratively exchanging the CEF-induced NCI signal and NCI-based AIDC power consumption.
    \item We develop a hierarchical CA-MARL framework based on a MAT method to prevent combinatorial action-space explosion of agents. In this framework, given the training/inference workloads spatially allocated by a global WM agent of the AIDC aggregator among AIDCs, each local AIDC agent of the AIDC operator executes the temporal training workload shifts and spatial GPU allocation for training/inference workloads while controlling the supply air temperature of the CRAC-based cooling system.
    \item We present a simple yet effective method, namely the largest remainder rule, to determine the actions of the WM and AIDC agents. Using this rule, their continuous actions for the spatial allocation of training/inference workloads at the AIDC level, as well as for the temporal shift of training workloads at the GPU level, are transformed into discrete ones to further reduce the dimensionality of the action space.
\end{enumerate}

The remainder of this paper is organized as follows. Sec.~\ref{sec:prelim} reviews the background of MAT. Sec.~\ref{sec:system} presents the models of the power distribution system, CEF, and AIDC. The proposed hierarchical CA-MARL framework is formulated in Sec.~\ref{sec:problem}. The corresponding simulation results are presented in Sec.~\ref{sec:performance}. Sec.~\ref{sec:conclusion} concludes this paper.

\section{Preliminaries} \label{sec:prelim}
In this section, we introduce MAT, which serves as the core learning architecture for the proposed hierarchical CA-MARL framework.

\BfPara{Multi-Agent Transformer} MAT \cite{Wen2022MAT} is adopted to model the cooperative yet heterogeneous interactions between the WM agent and local AIDC agents. 
MAT is grounded in the principle of \textit{multi-agent advantage decomposition}, which interprets joint decision-making as a sequential process in which each agent's action contributes an incremental advantage conditioned on previously selected actions \cite{kuba2021}. 
This perspective enables joint policy optimization without explicitly enumerating the exponentially large joint action space, while coordination is achieved through ordered conditioning rather than simultaneous action selection.
Building on this principle, MAT implements an actor–critic architecture using a transformer-based encoder–decoder structure. The shared encoder utilizes self-attention mechanisms to learn global representations of system states and capture complex inter-agent dependencies, whereas the masked autoregressive decoder generates the actions of the agents sequentially in an autoregressive manner. Rather than using a permanently fixed agent-specific order, the decoding order is randomly permuted during training iterations so that no particular agent is consistently assigned precedence.

The critic evaluates the encoded global states to estimate the value functions, whereas the actor updates the policies using an on-policy learning scheme based on proximal policy optimization (PPO) with a clipped surrogate objective.
This clipped objective prevents large, destructive policy updates, thereby ensuring a monotonic improvement. To stabilize training and reduce variance in policy updates, advantage estimates are computed using generalized advantage estimation (GAE) \cite{schulman2015gae}. 
This design enables each agent to enhance its local advantage while maintaining stable joint learning behavior. Following on-policy training, the trajectories collected from environment interactions are stored in a rollout buffer, which is used to compute returns and advantage estimates for PPO-based policy updates. For specific algorithm details, we refer the readers to \cite{MAT}. 

During training, agents access global information, including NCI signals and system-wide workload states, to learn coordinated policies, whereas each agent acts autonomously based on local observations during execution. Under the centralized-training and decentralized-execution setting, MAT is particularly suitable for the proposed carbon-aware AIDC scheduling problem due to its ability to jointly handle hierarchical coordination, strong inter-agent coupling, and action-space scalability. In the proposed framework, the WM agent determines the spatial allocation of heterogeneous workloads across geographically distributed AIDCs, and each local AIDC agent subsequently determines its execution strategy through temporal job shifting, GPU allocation, and cooling control. These decisions are hierarchically organized yet tightly coupled through shared grid conditions, NCI signals, workload backlogs, and local resource availability. Such characteristics make conventional MARL approaches less effective, as they either assume weak coupling among agents or rely on joint action representations that scale poorly with the number of agents and control variables. By contrast, MAT leverages a self-attention encoder to capture global inter-agent dependencies and an autoregressive decoder to model the conditional WM-to-AIDC decision flow, which reflects the hierarchical structure problem. Furthermore, MAT enables coordinated policy learning without explicitly enumerating the full joint action space, making it particularly suitable for large-scale AIDC scheduling.

\section{System model} \label{sec:system}

In this section, we present a system model for the operation of the proposed CA-MARL framework. It describes the operating conditions of the power distribution system (Sec.~\ref{subsec:distribution_system}), CEF-based carbon tracing (Sec.~\ref{subsec:carbon_tracing}), and AIDC with heterogeneous AI workloads and CRAC-based cooling system (Sec.~\ref{subsec:AIDC Model}).

\subsection{Power Distribution System Model}\label{subsec:distribution_system}

For nodes $i,j\in\mathcal{N}$ and lines $ij, ji\in\mathcal{L}$, which reflect the direction of the power flow, a convex power flow model is expressed as follows:
{\allowdisplaybreaks
\begin{align}
	P(Q)_{ji,t}&+P(Q)_{ij,t}+r(x)_{ij}\,\ell_{ij,t}=0, \label{eq:lineP}\\
	\sum_{j\in\mathcal{N}_i} P(Q)_{ji,t} &+ P(Q)^{\mathrm{inj}}_{i,t} = 0, \label{eq:busP}\end{align}
    \begin{multline}
	u_{i,t}-u_{j,t}	= -2\left(r_{ij}P_{ji,t}+x_{ij}\,Q_{ji,t}\right) \\
	                     - (r_{ij}^2+x_{ij}^2)\,\ell_{ij,t}, 
	                    \label{eq:vdrop}
    \end{multline}
    \begin{align}
    P^{\mathrm{inj}}_{i,t} &= P^{\mathrm{gt}}_{i,t} - \widehat{P}^\mathrm{load}_{i,t} - P^{\mathrm{AIDC}}_{i,t}, \label{eq:p_inj}\\
	Q^{\mathrm{inj}}_{i,t} &= Q^{\mathrm{gt}}_{i,t} - \widehat{Q}^\mathrm{load}_{i,t} - \eta P^{\mathrm{AIDC}}_{i,t}, \label{eq:q_inj}\\
	\ell_{ij,t}\,u_{j,t} \;&\ge\; P_{ij,t}^2 + Q_{ij,t}^2, \label{eq:ohm}\\
	\underline{V}^2 &\leq u_{i,t} \leq \overline{V}^2, \label{eq:voltagelimit}\\
	\underline{P}(\underline{Q})^\mathrm{gt}_{i} &\leq P(Q)^\mathrm{gt}_{i,t} \leq \overline{P}(\overline{Q})^\mathrm{gt}_{i}. \label{eq:gen_p_limit}
\end{align}}

The relationship between line losses and power flows is established in~\eqref{eq:lineP}, where the sum of two directional real (reactive) power flows ($P(Q)_{ij,t},P(Q)_{ji,t}$) balances the negative real (reactive) line loss ($r(x)_{ij}\ell_{ij,t}$). Here, $P(Q)_{ij,t}$ and $P(Q)_{ji,t}$ represent the real (reactive) power flows spanning from node $i$ to $j$ and from node $j$ to $i$ at time $t$, respectively; $r(x)_{ij}$ denotes the resistance (reactance) of line $ij$; and $\ell_{ij,t}$ indicates the squared current flowing from node $i$ to $j$.
Nodal power balance is enforced by~\eqref{eq:busP}, which requires that the net power flow from neighboring nodes $\mathcal{N}_i$ be equal to the real (reactive) power injection $P(Q)^{\mathrm{inj}}_{i,t}$ at node $i$ and time $t$.
The voltage drop between nodes $i$ and $j$ is modeled in~\eqref{eq:vdrop}, where $u_{i,t}$ is the squared voltage magnitude of node $i$.
The real (reactive) power injection $P(Q)^{\mathrm{inj}}_{i,t}$ defined in~\eqref{eq:busP} is described in~\eqref{eq:p_inj} and~\eqref{eq:q_inj}, expressed in terms of the gas-turbine (GT) real (reactive) power generation $P(Q)^{\mathrm{gt}}_{i,t}$, predicted real (reactive) power load demand $\widehat{P}(\widehat{Q})^\mathrm{load}_{i,t}$, and real power consumption $P^{\mathrm{AIDC}}_{i,t}$ of AIDC along with its corresponding fixed ratio of reactive to real power $\eta$.
The second-order cone constraint in~\eqref{eq:ohm} implements the relaxed convex form of Ohm's law. The squared voltage magnitude $u_{i,t}$ is bounded in~\eqref{eq:voltagelimit}, where $\underline{V}$ and $\overline{V}$ are the minimum
and maximum voltage magnitudes, respectively. The real (reactive) power generation of GT is limited in \eqref{eq:gen_p_limit}, where $\underline{P}(\underline{Q})^\mathrm{gt}_{i}$ and $\overline{P}(\overline{Q})^\mathrm{gt}_{i}$ are the minimum and maximum GT real (reactive) power outputs, respectively.

\subsection{Carbon Tracing Model Based on CEF}\label{subsec:carbon_tracing}

CEF is defined as a virtual network integrated into a power network to trace carbon footprints. Within this framework, a virtual carbon flow is defined as the theoretical flow of carbon emissions embedded in the active power flow from generation nodes to consumption nodes, allowing for granular carbon accounting \cite{CEF1}. A key concept of the CEF model is NCI $w_{i,t}$, which is defined as the ratio of total virtual carbon inflow $C^{\mathrm{inj}}_{i,t}$ [kgCO$_2$/t] to total real power inflow $P^{\mathrm{inj}}_{i,t}$ [kW] as follows:

\begin{equation}
w_{i,t} = \frac{C^{\mathrm{inj}}_{i,t}}{P^{\mathrm{inj}}_{i,t}} = \frac{w^\mathrm{gt}_{i,t} P^\mathrm{gt}_{i,t} + \sum_{h \in \mathcal{N}^+_i} w_{h,t} P_{hi,t}}{P^\mathrm{gt}_{i,t} + \sum_{h \in \mathcal{N}^+_i} P_{hi,t}},
\label{eq:CFE}
\end{equation}
where $P^\mathrm{gt}_{i,t}$ denotes the GT real power generation; $P_{hi,t}$ encodes the real power flow from node $h$ to $i$, where node $h$ belongs to a set $\mathcal{N}^+_i$ of neighboring nodes that send real power to node $i$; and $w^\mathrm{gt}_{i,t}$ represents the carbon emission factor of the GT.
Note that distribution-network losses are also considered in the CEF formulation. Following \cite{CEF1}, the carbon emissions associated with real power line losses are allocated based on the conservation of nodal carbon mass and the proportional sharing principle, where the line loss of each branch is assumed to have the same carbon intensity as that of the sending node. In addition, the CEF-integrated NCI represents a power-flow-aware nodal average carbon intensity, rather than a marginal carbon intensity. It reflects the carbon responsibility of electricity consumption at each node based on the physical power flows in the distribution network.

Based on NCI defined in \eqref{eq:CFE}, the virtual carbon flow equation is derived as follows:
\begin{align}
	w_{i,t}\!\left(P^\mathrm{gt}_{i,t} + \sum_{j\in\mathcal{N}_i} \hat p^{+}_{ji,t}\right)
	&= w^\mathrm{gt}_i P^\mathrm{gt}_{i,t} + \sum_{j\in\mathcal{N}_i} w_{j,t}\,\hat p^{+}_{ji,t}. \label{eq:cefbalance}
\end{align}

To offer a virtual carbon flow direction consistent with that
of the real power flow, a directed real power flow $P_{ij,t}$ is formulated as $P_{ij,t} = \hat p^{+}_{ij,t} - \hat p^{-}_{ij,t}$ along with the complementarity constraint $\hat p^{+}_{ij,t}\cdot\hat p^{-}_{ij,t} = 0$, where $\hat p^{+}_{ij,t}=\max\{P_{ij,t},0\}$ and $\hat p^{-}_{ij,t}=\max\{-P_{ij,t},0\}$. The non-convex nature of the complementarity constraint can be relaxed via linearization using the big-M method with binary decision variables, or a method based on special ordered set constraints.

Finally, while satisfying the constraints of power distribution system operation~\eqref{eq:lineP}--\eqref{eq:gen_p_limit} and CEF~\eqref{eq:cefbalance}, the CEF-integrated DSO problem aims to minimize the following augmented objective function, which comprises the total real power loss ($J_1$) and carbon emissions ($J_2$) of GTs and substation:
\begin{equation}
		\min\ J \;=\; \underbrace{\sum_{t\in\T}\sum_{ij\in\mathcal{L}} r_{ij}\,\ell_{ij,t}}_{J_1} \;+\; \lambda\,\underbrace{\sum_{t\in\T}\sum_{i\in\mathcal{N}^{\mathrm{gt}}\cup\mathcal{N}^{\mathrm{sub}}} w^\mathrm{gt}_i P^{\mathrm{gt}}_{i,t}}_{J_2} \label{eq:obj}
\end{equation}
where the non-negative $\lambda$ represents a penalty for carbon emissions. When $\lambda=0$, the DSO problem becomes a traditional problem without considering the reduction in carbon emissions.

\subsection{AIDC Model}\label{subsec:AIDC Model}

\subsubsection{Power Consumption Model}\label{sec:dc_power_model}

For each AIDC $i\in\mathcal{N}^{\text{AIDC}}$, the total power consumption comprises the aggregated server-based IT power $P^{\mathrm{IT}}_{i,t}$ and CRAC-based cooling power $P^{\mathrm{CRAC}}_{i,t}$:
\begin{equation}
P^{\mathrm{AIDC}}_{i,t}=P^{\mathrm{IT}}_{i,t}+P^{\mathrm{CRAC}}_{i,t},
\label{eq:dc-total}
\end{equation}
where $P^{\mathrm{IT}}_{i,t}$ in~\eqref{eq:dc-total} is decomposed into a fixed idle power $P^{\mathrm{idle}}_{i}$ and two workload-dependent powers:
\begin{equation}
P^{\mathrm{IT}}_{i,t}
= P^{\mathrm{idle}}_{i}
+\alpha^{\mathrm{train}}\sum_{m\in\mathcal M^{\mathrm{train}}} y^{\mathrm{train}}_{m,i,t}
+\alpha^{\mathrm{inf}}\sum_{m\in\mathcal M^{\mathrm{inf}}} y^{\mathrm{inf}}_{m,i,t},
\label{eq:dc-it}
\end{equation}
where $y^{\mathrm{train}}_{m,i,t}$ and $y^{\mathrm{inf}}_{m,i,t}$ are the throughputs of class $m$-based training and inference jobs [Tera operations per second, TOPS] with their corresponding power coefficients $\alpha^{\mathrm{train}}$ and $\alpha^{\mathrm{inf}}$ [kW/TOPS], respectively.

\begin{figure}[t!]
    \centering
    \includegraphics[width=0.95\linewidth]{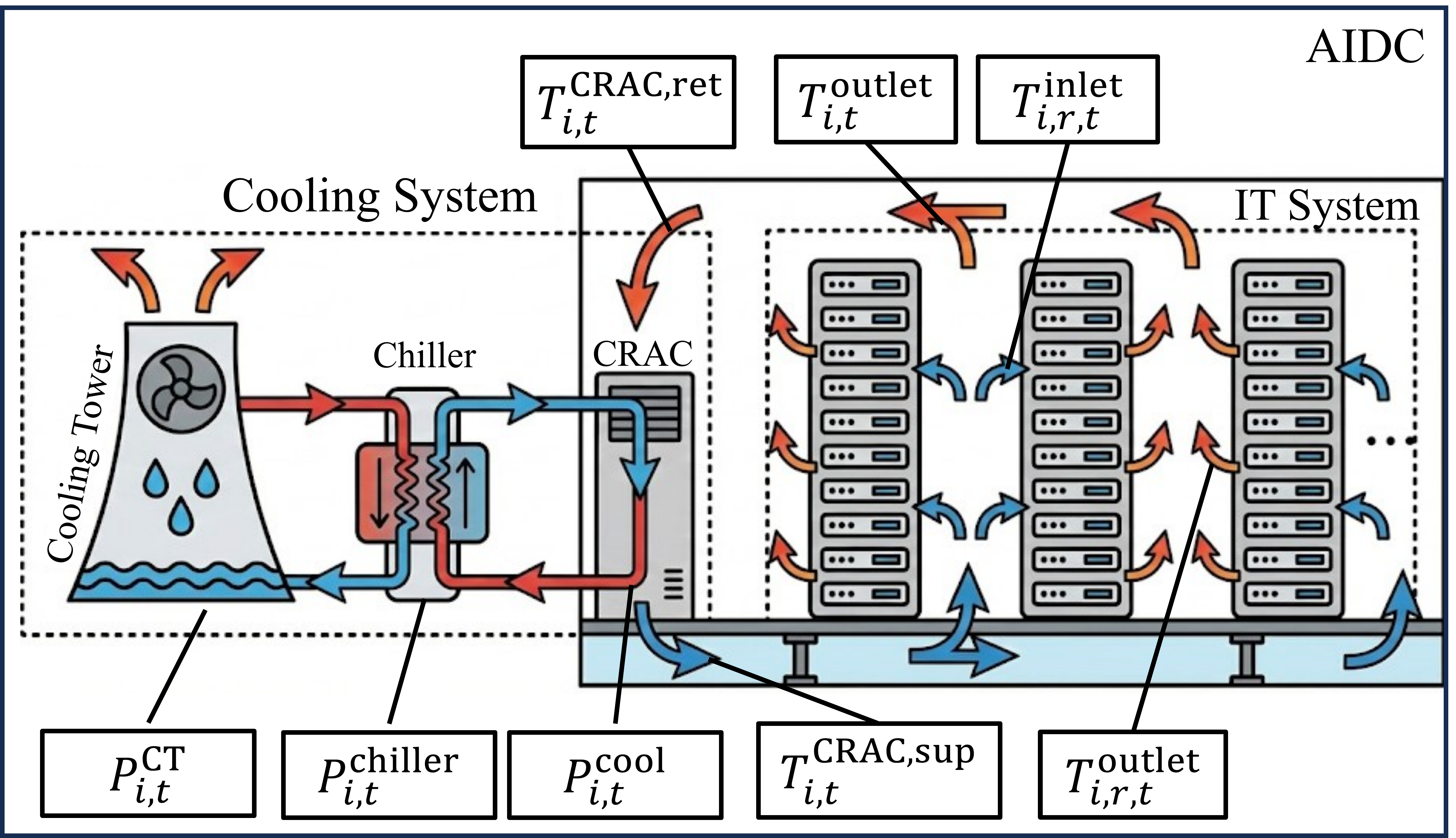}
    \caption{CRAC-based cooling system for AIDC.}
    \label{fig:cooling}

\end{figure}

\subsubsection{AI Workload Model}\label{sec:workload_model}
\begin{figure*}[ht!]
    \centering
    \includegraphics[width=0.99\linewidth]{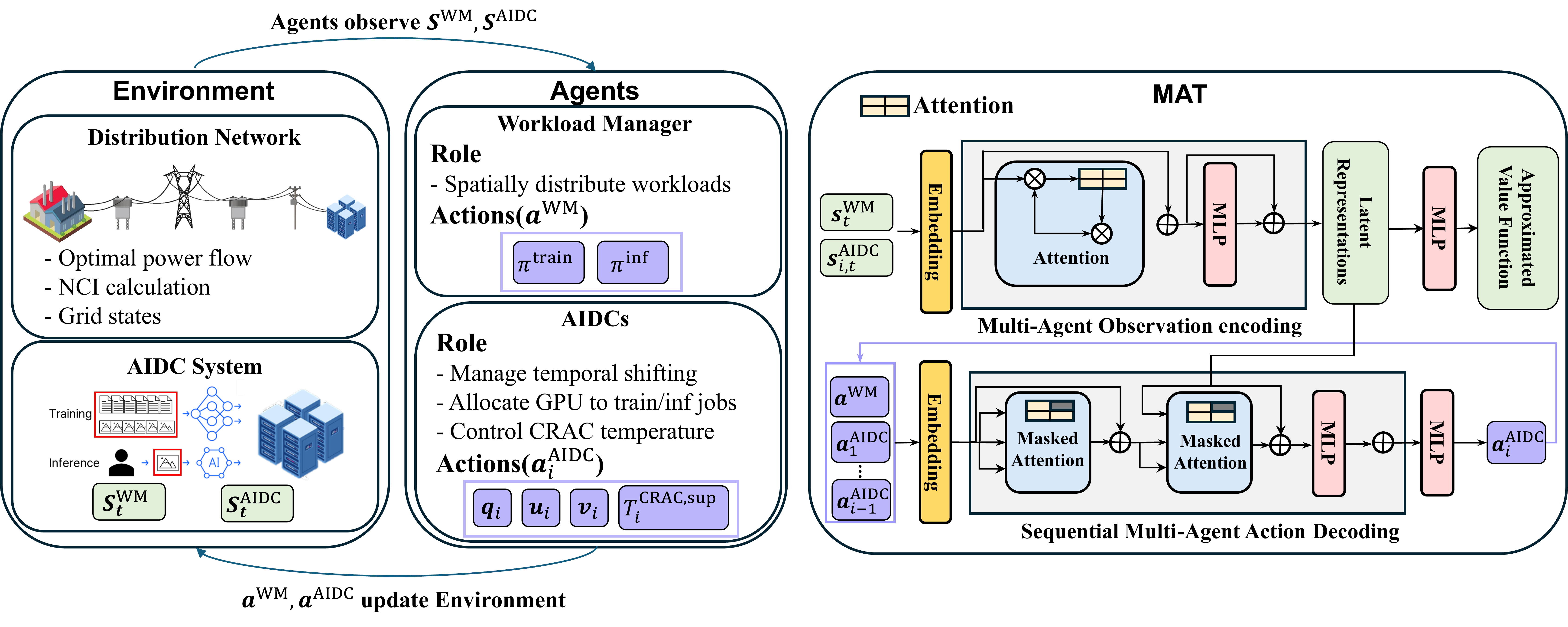}
    \caption{Structure of the proposed hierarchical CA-MARL framework comprising environment, agents, and MAT.}
    \label{fig:algo}

\end{figure*}
At each time slot $t$, heterogeneous AI workloads of type $k\in\{\mathrm{train},\mathrm{inf}\}$ and class $m\in\mathcal M^k$ arrive from external sources. The total number of AI workloads with type $k$ and class $m$ at time $t$ is denoted by $N^{k}_{m,t}\in\mathbb Z_{\ge0}$ (i.e., $N^{\mathrm{train}}_{m,t}:=N^{k}_{m,t}\big|_{k=\mathrm{train}}$ and $N^{\mathrm{inf}}_{m,t}:=N^{k}_{m,t}\big|_{k=\mathrm{inf}}$).

Given a fixed computational size $j^{k}_{m}$ [Tera operations (TO)/job] along with a scheduling resolution $\Delta t$ [min], the training (inference) throughput in~\eqref{eq:dc-it} is defined as $y^{\mathrm{train(inf)}}_{m,i,t}
= j^{\mathrm{train(inf)}}_{m}\,x^{\mathrm{train(inf)}}_{m,i,t}/(\Delta t \times 60)$.
Here, $x^{\mathrm{train(inf)}}_{m,i,t}$ is the total number of executed class $m$-based training (inference) jobs for AIDC $i$ at time $t$.
Moreover, $x^{\mathrm{train(inf)}}_{m,i,t}$ is associated with the number of allocated GPUs for training (inference), denoted as $g^{\mathrm{train(inf)}}_{i,t}$ (i.e., action-related variables of AIDC agent in Sec.~\ref{sec:problem}); it satisfies the following constraint:
\begin{align}
\sum_{m\in\mathcal M^{\mathrm{train(inf)}}}
j^{\mathrm{train(inf)}}_{m}\,x^{\mathrm{train(inf)}}_{m,i,t}
&= \xi^{\mathrm{train(inf)}}_{\max}\,g^{\mathrm{train(inf)}}_{i,t}\,\Delta t,
\label{eq:train-capacity}
\end{align}
where $\xi^{\mathrm{train(inf)}}_{\max}$ [TO/GPU/min] denotes the per-GPU peak processing rate for training (inference).

\BfPara{Training Jobs}
For AIDC $i$, each delay-tolerant training job in class $m\in\mathcal M^{\mathrm{train}}$ has a deadline $\tau^{\mathrm{train}}_m$ with deferral $h$, which is selected from a set of allowable deferral steps $\Delta_m \subseteq \{0,\dots,\tau^{\mathrm{train}}_m\}$. 
The dynamics of bucket $B^{h+1}_{m,i,t}$ containing class-$m$ training backlogs with deferral step $h+1\in\Delta_m$ is expressed as follows:
\begin{align}
B^{h}_{m,i,t+1}
&= B^{h+1}_{m,i,t} + n^{h+1}_{m,i,t},
\quad h = 0,\dots,\tau^{\mathrm{train}}_m - 1,
\label{eq:train-backlog-roll}\\
B^{\tau^{\mathrm{train}}_m}_{m,i,t+1}
&= 0,
\label{eq:train-backlog-reset}
\end{align}
where $n^{h+1}_{m,i,t}$ represents the number of newly arriving class-$m$ training jobs that must be executed after $h+1$ time steps from time $t$. Note that the value of $n^{h+1}_{m,i,t}$ is determined by the time shifting action of the training job for the AIDC agent described in Sec.~\ref{subsec:AIDC_Agent}. Let us define $D_{m,i,t}$ as $D_{m,i,t}=B^{0}_{m,i,t}+n^{0}_{m,i,t}$, expressing the number of non-deferrable training backlogs with $h=0$ at current time $t$.
The number of dropped training jobs is then defined as follows:
\begin{equation}
l^{\mathrm{train}}_{m,i,t}=\max\left(0,D_{m,i,t}-x^{\mathrm{train}}_{m,i,t}\right).
\label{eq:train-deadline}
\end{equation} 

\BfPara{Inference Jobs} Each delay-sensitive inference job follows a first-in, first-out queueing model, which is described as follows:
\begin{equation}
Z_{m,i,t+1}
=\max\left(0,\,
Z_{m,i,t}+n^{\mathrm{inf}}_{m,i,t}
- x^{\mathrm{inf}}_{m,i,t}
- l^{\mathrm{inf}}_{m,i,t}\right),
\label{eq:inf-queue}
\end{equation}
where $Z_{m,i,t}$ is the queue length of class-$m$ inference backlogs for AIDC $i$ at time $t$. $n^{\mathrm{inf}}_{m,i,t}$ and  $x^{\mathrm{inf}}_{m,i,t}$ are the number of allocated and executed inference jobs, respectively, which are computed by the actions of WM and AIDC agents mentioned in Sec.~\ref{subsec: WM Agent} and Sec.~\ref{subsec:AIDC_Agent}, respectively. $l^{\mathrm{inf}}_{m,i,t}$ is the number of inference jobs that are dropped when their waiting times reach the deadline $\tau^{\mathrm{inf}}_m$.

\subsubsection{CRAC-Based Cooling Model}

As shown in Fig.~\ref{fig:cooling}, a cooling system for AIDC comprises a CRAC unit, a chiller, and a cooling tower. The mathematical formulations of the thermal dynamics and power consumption of these components are provided in~\cite{sustaindc}. The rack inlet temperature $T^{\mathrm{inlet}}_{i,r,t}$ of rack $r\in\mathcal{R}_i$ in AIDC $i$ at time $t$ equals the sum of the spatial temperature offset $\Delta T^{\mathrm{sup}}_{i,r}$ and CRAC supply air temperature $T^{\mathrm{CRAC,sup}}_{i,t}$:
\begin{equation}
    T^{\mathrm{inlet}}_{i,r,t} = \Delta T^{\mathrm{sup}}_{i,r} + T^{\mathrm{CRAC,sup}}_{i,t}.
\label{eq:CRAC_sup}
\end{equation}
Based on the IT power consumption $P^{\mathrm{IT}}_{i,t}$, the average outlet temperature $T^{\mathrm{outlet}}_{i,t}$ and CRAC return temperature $T^{\mathrm{CRAC,ret}}_{i,t}$ are expressed as follows:
\begin{align}
    T^{\mathrm{outlet}}_{i,t} &= \frac{1}{|\mathcal{R}_i|}\sum_{r\in\mathcal{R}_i}
    \left(T^{\mathrm{inlet}}_{i,r,t} + \frac{P^{\mathrm{IT}}_{i,t}}{c^{\mathrm{air}}\rho^{\mathrm{air}}\Phi^{\mathrm{sfan}}}\right), \label{eq:temp_outlet}\\
    T^{\mathrm{CRAC,ret}}_{i,t} &= \frac{1}{|\mathcal{R}_i|}\sum_{r\in\mathcal{R}_i}
    \left(\Delta T^{\mathrm{CRAC,ret}}_{i,r} + T^{\mathrm{outlet}}_{i,t}\right), \label{eq:temp_return}
\end{align}
where $c^{\mathrm{air}}$ and $\rho^{\mathrm{air}}$ are the air thermal capacity [J/(kg$\cdot$°C)] and density [kg/m$^3$], respectively; $\Phi^{\mathrm{sfan}}$ is the fan airflow rate [m$^3$/s]; and $\Delta T^{\mathrm{CRAC,ret}}_{i,r}$ is the rack-specific return approach temperature.
Using the cooling load $P^{\mathrm{cool}}_{i,t}$ calculated by $P^{\mathrm{cool}}_{i,t} = \rho^{\mathrm{air}}\Phi^{\mathrm{sfan}} c^{\mathrm{air}} (T^{\mathrm{CRAC,ret}}_{i,t}-T^{\mathrm{CRAC,sup}}_{i,t})$, the power consumptions of the chiller ($P^{\mathrm{chiller}}_{i,t}$) and cooling tower ($P^{\mathrm{CT}}_{i,t}$) are respectively expressed as follows:
{\allowdisplaybreaks
\begin{align}
    P^{\mathrm{chiller}}_{i,t} &= \frac{P^{\mathrm{cool}}_{i,t}}{\operatorname{COP}_i(T^{\mathrm{CRAC,sup}}_{i,t})}, \label{eq:chiller}\\
    P^{\mathrm{CT}}_{i,t} &= P^{\mathrm{CT,REF}}_i 
    \left(\frac{P^{\mathrm{chiller}}_{i,t}}{c^{\mathrm{air}}\rho^{\mathrm{air}}\Delta T^{\mathrm{CT,air}}_{i,t}\Phi^{\mathrm{CT,REF}}_i}\right)^3, \label{eq:cooling_tower}
\end{align}
}
where $\operatorname{COP}_i(\cdot)$ is the coefficient of performance (COP); $P^{\mathrm{CT,REF}}_i$ and $\Phi^{\mathrm{CT,REF}}_i$ represent the reference fan power and airflow rate of the cooling tower, respectively; and $\Delta T^{\mathrm{CT,air}}_{i,t}$ denotes the rise of the cooling air temperature. Heat generated by the IT system increases the air temperature from the rack inlet temperature $T^{\mathrm{inlet}}_{i,r,t}$ to the rack outlet temperature $T^{\mathrm{outlet}}_{i,r,t}$, and is delivered to the CRAC as return air with temperature $T^{\mathrm{CRAC,ret}}_{i,t}$.
The extracted heat is then removed through the cooling-water loop (CRAC-chiller-cooling tower), which forms a closed-loop heat-flow path of the system.

Finally, the CRAC-based cooling system consumes a total power expressed as $P^{\mathrm{CRAC}}_{i,t}=P^{\mathrm{chiller}}_{i,t} + P^{\mathrm{CT}}_{i,t}$, while satisfying the following temperature limits:
\begin{equation}
\begin{aligned}
    T^{\mathrm{inlet}}_{i,r,t}
    \le &\overline{T}^{\mathrm{inlet}}_{i,r},\\
   \underline{T}_i^{\mathrm{CRAC,sup}} \le &T^{\mathrm{CRAC,sup}}_{i,t}
    \le \overline{T}_i^{\mathrm{CRAC,sup}}.
\label{eq:temp_inequal}
\end{aligned}
\end{equation}

Note that the CRAC-based cooling model ensures thermal safety in terms of $T^{\mathrm{inlet}}_{i,r,t}$ and $T^{\mathrm{CRAC,sup}}_{i,t}$. The outlet temperature $T^{\mathrm{outlet}}_{i,t}$ and CRAC return temperature $T^{\mathrm{CRAC,ret}}_{i,t}$ are algebraically determined by the inlet temperature and bounded IT power consumption, and therefore remain implicitly within safe operating limits without requiring additional constraints.

\section{Proposed Hierarchical CA-MARL Framework}\label{sec:problem}

In this section, we present the proposed hierarchical CA-MARL framework for economical and eco-friendly DSO--AIDC coordination based on the system model described in Sec.~\ref {sec:system}. As shown in Fig.~\ref{fig:algo}, the proposed framework includes the following three key components. 

\noindent\textbf{Environment}: It encompasses the CEF-integrated DSO problem (Sec.~\ref{subsec:distribution_system} and Sec.~\ref{subsec:carbon_tracing}) and AIDC operational dynamics (Sec.~\ref{subsec:AIDC Model}). The former determines the operating conditions of the CEF-integrated power distribution system and calculates the NCI and AIDC power consumption. The latter classifies and processes the arriving training/inference jobs for AIDC while monitoring the condition of CRAC-based cooling.

\noindent\textbf{Agents}: They are classified into two types following a hierarchical structure: i) WM agent (global layer) and ii) AIDC agent (local layer). Both agents interact with the aforementioned environment to identify their states. The WM agent spatially distributes training/inference jobs to all AIDCs. Based on the distributed jobs, each AIDC agent performs a temporal shift of training jobs, allocates training GPU blocks and inference GPUs, and controls the CRAC supply air temperature. Finally, the AIDC power consumption calculated by each AIDC agent is fed back into the constraint of the DSO problem in the environment. The interaction between the DSO and the agents (i.e., the WM and AIDC agents) is implemented as a sequential closed-loop mechanism with time-scale separation, where the DSO updates the NCI signals at a slower time scale.
Within each update interval, the NCI signals are treated as piecewise-constant inputs, which ensures a temporally consistent environment and stable coordination between the DSO and the agents. The hierarchical structure consisting of the WM and AIDC agents,  together with well-designed state, action, and reward formulations, improves the stability and tractability of the learning process. The state/action space and reward function of the WM and AIDC agents are described in subsequent sections.

\noindent\textbf{MAT}: It implements the proposed hierarchical coordination via the multi-agent advantage decomposition theorem with an autoregressive execution order tailored to our system structure, in which the WM agent always acts first and the relative order of the AIDC agents is randomly permuted during training, i.e., $i_{1:n} = (\text{WM}, \text{AIDC}_{\sigma(1)}, \dots, \text{AIDC}_{\sigma(|\mathcal{N}^{\text{AIDC}}|)})$, where $\sigma$ denotes a permutation of the AIDC indices. As outlined in Algorithm~\ref{alg:hierarchical-mat}, MAT implements this ordering through an encoder-decoder architecture: conditioned on the global representation from the shared encoder, the WM agent first generates the spatial placement decision $\boldsymbol a^{\mathrm{WM}}_t$ (\textit{where} to place jobs), which is then provided as conditioning information to the local AIDC agents. Subsequently, each AIDC agent $i$ produces its local decision $\boldsymbol a^{\mathrm{AIDC}}_{i,t}$, determining \textit{when} and \textit{how} to execute the jobs (temporal deferral, GPU allocation, and cooling control). This explicit WM$\rightarrow$AIDC autoregressive decision flow prevents combinatorial action-space explosion while preserving coordination under centralized training with decentralized execution. For conciseness, we omit repeated definitions and refer to the detailed state/action/reward specification and task-wise execution using the corresponding section and equation indices.

\begin{algorithm}[t!]
\caption{Hierarchical CA-MARL Training via MAT}
\small
\label{alg:hierarchical-mat}
\begin{algorithmic}[1]
\STATE Initialize MAT: shared encoder, decoder (actor), and critic networks; initialize an on-policy rollout buffer.
\FOR{episode $=1,2,\dots$}
    \STATE Reset environment; clear the rollout buffer.
    \FOR{$t=0,1,\dots,T-1$}
        \STATE Update environment and observe WM state $\boldsymbol s^{\mathrm{WM}}_t$ by \eqref{eq:state_wm_agent}.
        \STATE \textbf{Encoding:} Generate global representations from joint observations via the shared encoder.
        \STATE \textbf{Decoding (WM):} Sample WM action $\boldsymbol a^{\mathrm{WM}}_t$ via the decoder conditioned on encoder outputs.
        \STATE Obtain integer job allocations $\{n^{k}_{m,i,t}\}$ via the largest remainder rule (Sec.~\ref{subsec: WM Agent}).
        \FOR{each AIDC $i\in\mathcal N^{\text{AIDC}}$ in the MAT order}
            \STATE Observe AIDC state $\boldsymbol s^{\mathrm{AIDC}}_{i,t}$ by \eqref{eq:state_aidc_agent}.
            \STATE \textbf{Decoding (AIDC):} Sample AIDC action $\boldsymbol a^{\mathrm{AIDC}}_{i,t}$ autoregressively conditioned on encoder outputs and preceding actions $(\boldsymbol a^{\mathrm{WM}}_t, \boldsymbol a^{\mathrm{AIDC}}_{1:i-1,t})$.
            \STATE Execute \textbf{Tasks 1)--4)} (Sec.~\ref{subsec:AIDC_Agent}) to update local workloads and queues.
            \STATE Calculate $P^{\mathrm{AIDC}}_{i,t}$ by \eqref{eq:dc-total}--\eqref{eq:dc-it}.
        \ENDFOR 
        \STATE Execute joint actions in the environment to transition state $s_{t}$ to $s_{t+1}$; compute reward $\texttt{R}_t$ by \eqref{eq:global-reward}.
        \STATE Store the transition in the rollout buffer.
    \ENDFOR
    \STATE Estimate returns/advantages using GAE from the rollout buffer.
    \STATE Update MAT value (critic) and policy (actor) networks using clipped PPO with the collected rollouts.
\ENDFOR
\end{algorithmic}
\end{algorithm}
\subsection{WM Agent}\label{subsec: WM Agent}

\begin{figure}[t!]
    \centering
    \includegraphics[width=0.99\linewidth]{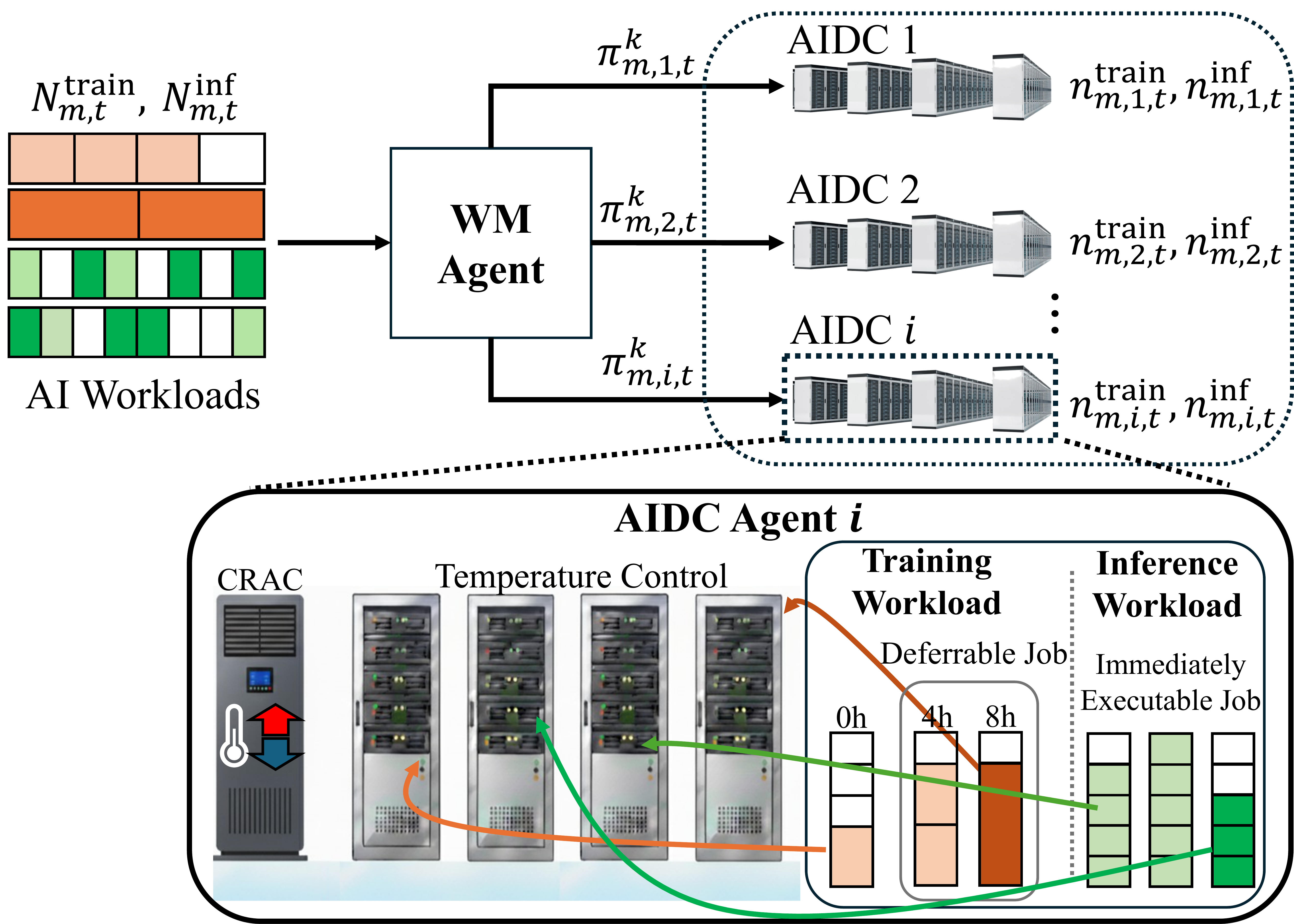}
    \caption{Spatio-temporal workload scheduling of WM and AIDC agents.}
    \label{fig:workload_manager}

\end{figure}

The WM agent aims to spatially allocate optimal numbers of the training and inference jobs ($n^{\mathrm{train}}_{m,i,t},n^{\mathrm{inf}}_{m,i,t}$) into each AIDC $i$ based on a fractional allocation ratio $\pi^{k}_{m,i,t}$ during the arrivals of the requested training and inference jobs ($N^{\mathrm{train}}_{m,t},N^{\mathrm{inf}}_{m,t}$) as shown in Fig.~\ref{fig:workload_manager}.

\noindent\textbf{State Space:} The state of the environment is defined as:
\begin{align}
\boldsymbol s^{\mathrm{WM}}_t
=\Big[
& \kappa_t, \boldsymbol w_{t},\boldsymbol S^{\mathrm{AIDC}}_{t},
\boldsymbol N^{\mathrm{train}}_{t}, 
\boldsymbol N^{\mathrm{inf}}_{t},\
t 
\Big].\label{eq:state_wm_agent}
\end{align}

Here, $\kappa_t$ is a time-of-use (TOU) electricity price at time $t$ and $\boldsymbol w_{t}$ is the vector of NCI values for all AIDCs at time $t$. Note that $\boldsymbol w_{t}$ is calculated and obtained from the CEF-integrated DSO problem described in Secs.~\ref{subsec:distribution_system} and~\ref{subsec:carbon_tracing}, which serves as a critical signal that induces the WM agent to allocate training and inference jobs for low-carbon operation of AIDC. $\boldsymbol N^{\mathrm{train}}_{t}$ and $\boldsymbol N^{\mathrm{inf}}_{t}$ are the vectors of the total numbers of arriving training and inference jobs, respectively, for all classes $m\in \mathcal M^{\mathrm{train}}\cup \mathcal M^{\mathrm{inf}}$ at time $t$. 
The vector $\boldsymbol S^{\mathrm{AIDC}}_{t}$ is defined as $\boldsymbol S^{\mathrm{AIDC}}_{i,t}=\left[\bar B_{i,t},\,\bar Z_{i,t},\,P^{\mathrm{AIDC}}_{i,t},\,\phi^{\mathrm{free}}_{i,t}\right]$ for all AIDCs $i\in\mathcal{N}^{\text{AIDC}}$, where $\bar B_{i,t}$ and $\bar Z_{i,t}$ are the total numbers of training backlogs in all buckets and inference backlogs in all queues at time $t$, respectively. $P^{\mathrm{AIDC}}_{i,t}$ denotes the power consumption of AIDC at time $t$. $\phi^{\mathrm{free}}_{i,t}\in [0,1]$ denotes the available GPU capacity ratio at AIDC $i$. It is calculated as $1 - (g^{\mathrm{train}}_{i,t-1} + g^{\mathrm{inf}}_{i,t-1})/G_i$, where $g^{\mathrm{train(inf)}}_{i,t-1}$ is the number of GPUs used for training (inference) at previous time step $t-1$ and $G_i$ is the GPU capacity.

\noindent\textbf{Action Space:} The action of the WM agent is defined as:
\begin{align}
\boldsymbol a^{\mathrm{WM}}_t
=\left[\pi^{k}_{m,i,t}\right]~~
\text{s.t.}~\sum_{i\in\mathcal N^{\text{AIDC},k}_{m}} \pi^{k}_{m,i,t}=1,~\pi^{k}_{m,i,t}\ge0,
\end{align}
where $\pi^{k}_{m,i,t}$ is a normalized spatial distribution ratio of heterogeneous training jobs for AIDC $i$ in type $k$ and class $m$.
The value of $\pi^{k}_{m,i,t}$ is used to calculate the spatial allocation of training jobs for all AIDCs via the largest remainder rule in the following steps.
First, the non-integer spatial allocation of training jobs $\hat n^{k}_{m,i,t}=\pi^{k}_{m,i,t} N^k_{m,t}$ is calculated based on the value of $\pi^{k}_{m,i,t}$.
Second, the decimal part of $\hat n^{k}_{m,i,t}$ is defined as $f^{k}_{m,i,t}=\hat n^{k}_{m,i,t}-\ddot n^{k}_{m,i,t}$, which is computed by flooring $\hat n^{k}_{m,i,t}$ to $\ddot n^{k}_{m,i,t}=\lfloor \hat n^{k}_{m,i,t}\rfloor$.
Third, the remainder is calculated as $R^k_{m,t}=N^k_{m,t}-\sum_{i\in\mathcal N_{m}^{\text{AIDC},k}} \ddot n^{k}_{m,i,t}$. 
Finally, for all $q=1,\dots,|\mathcal N_{m}^{\text{AIDC},k}|$, a permutation of AIDCs in $\mathcal N_{m}^{\text{AIDC},k}$ is expressed using $\sigma(q)$ and the decimal parts are reordered according to the largest remainder rule as follows: $f^{k}_{m,\sigma(1),t}\ge\cdots \ge f^{k}_{m,\sigma(|\mathcal N_{m}^{\text{AIDC},k}|),t}$.
Subsequently, the integer value of the spatial allocation of the training jobs is obtained using $n^{k}_{m,\sigma(q),t}=\ddot n^{k}_{m,\sigma(q),t}+\mathbf 1\!\{q\le R^k_{m,t}\}$ where $\mathbf 1\{\cdot\}$ represents an indicator, which equals $1$ if $q\le R^k_{m,t}$; otherwise it equals $0$. Note that the continuous action is used as a relaxed parameterization of a discrete job allocation problem. The integer allocation obtained via the largest remainder rule deviates from the continuous ratio allocation by less than one job for each allocation variable, thereby strictly limiting the quantization error while preserving the total number of jobs. In addition, during training, the continuous policy output is immediately projected into a feasible integer allocation before interacting with the environment. Consequently, the environment transition and reward are determined by the executed discrete action rather than the intermediate continuous ratio output, thereby ensuring consistency between training and execution.

\subsection{AIDC Agent}\label{subsec:AIDC_Agent}

For AIDC $i$, the AIDC agent schedules an optimal AIDC operation by performing the following four tasks as shown in Fig.~\ref{fig:workload_manager}: \textbf{Task 1)} temporal shifting of training jobs; \textbf{Task 2)} allocation of training GPU blocks; \textbf{Task 3)} allocation of inference GPUs; and \textbf{Task 4)} control of CRAC supply air temperature.

\noindent\textbf{State Space:} The state of each AIDC $i$ agent is defined as:
\begin{align}
\boldsymbol s^{\mathrm{AIDC}}_{i,t}
=\Big[ \kappa_t,\boldsymbol\kappa_{t+h},\boldsymbol B_{i,t},w_{i,t},P^{\mathrm{AIDC}}_{i,t-1},\phi^{\mathrm{free}}_{i,t},\boldsymbol n^{\mathrm{train}}_{i,t},\boldsymbol n^{\mathrm{inf}}_{i,t},t \Big],
\label{eq:state_aidc_agent}
\end{align}
where $\kappa_t$ is the TOU electricity price at time $t$; $\boldsymbol\kappa_{t+h}$ is the vector of electricity prices with the deferral time slot $h\in \Delta_m$; $\boldsymbol B_{i,t}$ is the vector of total numbers of training backlogs for all classes $m\in \mathcal M^{\mathrm{train}}$ at time $t$; $w_{i,t}$ is the NCI value at time $t$; $P^{\mathrm{AIDC}}_{i,t-1}$ is the AIDC power consumption at the previous time $t-1$, and $\phi^{\mathrm{free}}_{i,t}$ represents the ratio of available GPU capacity as defined in Sec.~\ref{subsec: WM Agent}. In addition, $\boldsymbol n^{\mathrm{train}}_{i,t}$ and $\boldsymbol n^{\mathrm{inf}}_{i,t}$ are the vectors of numbers of training and inference jobs, respectively, updated by the action of the WM agent (described in Sec.~\ref{subsec: WM Agent}).

\noindent\textbf{Action Space:} The action of the AIDC agent, which is associated with the four aforementioned tasks, is defined as:
\begin{align}
\boldsymbol a^{\mathrm{AIDC}}_{i,t}
=\Big[
\boldsymbol q_{i,t},\boldsymbol u_{i,t},\boldsymbol v_{i,t},T^{\mathrm{CRAC,sup}}_{i,t}\Big].
\label{eq:AIDC action}
\end{align}

\noindent\textbf{Task 1):} $\boldsymbol q_{i,t}=\left[q^h_{m,i,t}\right]$ is the vector of normalized non-negative temporal shifting ratios of training jobs for any class $m$ in a deferral time slot $h\in\Delta_m\subseteq\{0,\dots,\tau^{\mathrm{train}}_m\}$ while satisfying $\sum_{h\in\Delta_m} q^h_{m,i,t}=1$.
This action shifts a fraction of training jobs from the current to the future.
For example, $q^4_{m,i,t}=0.3$ indicates that $30\%$ of class $m$-related training jobs for AIDC $i$ at current time $t$ are processed after four time steps ($h=4$).
The largest remainder rule, applied to calculate the solution of the WM agent, is also employed using the values of $\boldsymbol q_{i,t}$ to calculate the temporal shifts in the training job allocation for all AIDCs, as explained next. 
First, we calculate the non-integer temporal shifting of training jobs $\hat n^{h}_{m,i,t}=q^h_{m,i,t}n^{\mathrm{train}}_{m,i,t}$, its flooring $\ddot n^{h}_{m,i,t}:=\lfloor\hat n^{h}_{m,i,t}\rfloor$, and its decimal parts $f^{h}_{m,i,t}=\hat n^{h}_{m,i,t}-\ddot n^{h}_{m,i,t}$. The remainder is then calculated as $R^{\mathrm{def}}_{m,i,t}
        = n^{\mathrm{train}}_{m,i,t}
           - \sum_{h\in\Delta_m}\ddot n^{h}_{m,i,t}$.
For all $p=1,\dots,|\Delta_m|$ with a permutation $\sigma^{\mathrm{def}}_{m,i,t}(p)$ of $\Delta_m$, the decimal parts are reordered based on the largest remainder rule as follows: $f^{\sigma^{\mathrm{def}}(1)}_{m,i,t}\ge \cdots\ge f^{\sigma^{\mathrm{def}}(|\Delta_m|)}_{m,i,t}$. Finally, the integer value of the temporal shifting of the training jobs is calculated as $n^{\sigma^{\mathrm{def}}(p)}_{m,i,t}
=\ddot n^{\sigma^{\mathrm{def}}(p)}_{m,i,t}
+\mathbf 1\!\{\,p\le R^{\mathrm{def}}_{m,i,t}\,\}$ while ensuring $\sum_{h=0}^{\tau^\mathrm{train}_m} n^{h}_{m,i,t}=n^{\mathrm{train}}_{m,i,t}$.

\noindent\textbf{Tasks 2) and 3):} $\boldsymbol u_{i,t}=[u_{m,i,t}]$ and $\boldsymbol v_{i,t}=[v_{m,i,t}]$ are the spatial allocation ratios for
training and inference jobs, respectively; $u_{m,i,t},v_{m,i,t}\in[0,1]$ specifies the fraction of available GPUs for class $m$-based training and inference jobs, while satisfying the following constraint:
\begin{equation}
\sum_{m\in\mathcal M^{\mathrm{train}}} u_{m,i,t}
+
\sum_{m\in\mathcal M^{\mathrm{inf}}} v_{m,i,t}
\le 1.
\label{eq:gpu-ratio-constraint}
\end{equation}

Using the values of $u_{m,i,t}$ and $v_{m,i,t}$, the total number of GPUs used for training and inference can be respectively expressed as follows:
\begin{align}
g^{\mathrm{train}}_{m,i,t}
&= b_m\Big\lfloor \frac{u_{m,i,t}\,G_i}{b_m}\Big\rfloor,
\quad m\in\mathcal M^{\mathrm{train}},\\
g^{\mathrm{inf}}_{m,i,t}
&= \Big\lfloor v_{m,i,t}\,G_i\Big\rfloor,
\quad m\in\mathcal M^{\mathrm{inf}},
\end{align}
where $b_m$ is the GPU block size for the class $m$-based training job.
The total number of these GPUs is limited by the GPU capacity $G_i$ (i.e., $\sum_{m\in\mathcal M^{\mathrm{train}}}g^{\mathrm{train}}_{m,i,t}+\sum_{m\in\mathcal M^{\mathrm{inf}}}g^{\mathrm{inf}}_{m,i,t}\le G_i$). 

\noindent\textbf{Task 4):} $T^{\mathrm{CRAC,sup}}_{i,t}$ is the supply air temperature of CRAC for AIDC $i$ at time $t$. This action is calculated using the constraints~\eqref{eq:CRAC_sup}--\eqref{eq:temp_inequal} of the CRAC cooling model described in Sec.~\ref{subsec:AIDC Model}.

\subsection{Reward Function}\label{subsec:reward}

Let us consider a situation in which the WM and AIDC agents share a single global reward computed by the environment. The reward $\texttt{R}_t$ of these agents is defined as a weighted multi-reward function with non-negative weights $c_1\sim c_5$. Specifically, it comprises one reward component for the total training and inference throughputs ($\texttt{R}_{1,t}$) and four penalty components: total electricity purchase cost ($\texttt{R}_{2,t}$), total carbon emissions associated with electricity consumption ($\texttt{R}_{3,t}$), and total number of dropped training and inference jobs ($\texttt{R}_{4,t},\texttt{R}_{5,t}$), expressed as follows:
\begin{align}
\texttt{R}_t
={}&c_1\!\underbrace{\left(\sum_{i\in\mathcal N^{\text{AIDC}}}\sum_{m\in\mathcal M^{\mathrm{train}}} y^{\mathrm{train}}_{m,i,t} + \sum_{i\in\mathcal N^{\text{AIDC}}}\sum_{m\in\mathcal M^{\mathrm{inf}}} y^{\mathrm{inf}}_{m,i,t}\right)}_{\texttt{R}_{1,t}}\nonumber
\label{eq:global-reward}\\
&-\,c_2\!\underbrace{\sum_{i\in\mathcal N^{\text{AIDC}}}\! \kappa_t\, P^{\mathrm{AIDC}}_{i,t}\,}_{\texttt{R}_{2,t}}-\,c_3\!\underbrace{\sum_{i\in\mathcal N^{\text{AIDC}}}\! w_{i,t}\, P^{\mathrm{AIDC}}_{i,t}\,\,}_{\texttt{R}_{3,t}}\\
&-\,c_4\!\underbrace{\sum_{i\in\mathcal N^{\text{AIDC}}}\!\sum_{m\in\mathcal M^{\mathrm{train}}}\! l^{\mathrm{train}}_{m,i,t}}_{\texttt{R}_{4,t}}-\,c_5\!\underbrace{\sum_{i\in\mathcal N^{\text{AIDC}}}\!\sum_{m\in\mathcal M^{\mathrm{inf}}}\! l^{\mathrm{inf}}_{m,i,t}}_{\texttt{R}_{5,t}}.\nonumber
\end{align}

The weights $c_1\sim c_5$ in \eqref{eq:global-reward} can be empirically selected through preliminary experiments, rather than through an exhaustive optimization procedure, to balance heterogeneous reward and penalty components and ensure stable learning process. Among them, $c_2$ and $c_3$ are the most influential weights because they directly determine the economic-environmental trade-off between electricity purchase cost and carbon emissions.

\subsection{Complexity Analysis of an Action-Space Cardinality}\label{subsec:action_complexity}
To quantify the mitigation of combinatorial action-space explosion through the proposed hierarchical CA-MARL framework, we compare the cardinality of the action space between a non-hierarchical (flat) centralized structure and the proposed hierarchical WM--AIDC structure. In the flat architecture, a single agent jointly performs the inter-AIDC workload allocation and the scheduling of all local AIDCs. By contrast, in the proposed hierarchical architecture, all decisions are decomposed into the policy of WM agent and the policies of $N$ local AIDC policies, where $N:=|\mathcal N^{\mathrm{AIDC}}|$.
For a fair comparison of the action-space cardinality between the flat and hierarchical structures, we discretize the ratio-type continuous actions ($\boldsymbol a^{\mathrm{WM}}_t,\boldsymbol q_{i,t},\boldsymbol u_{i,t},\boldsymbol v_{i,t}$) and the continuous CRAC supply temperature ($T^{\mathrm{CRAC,sup}}_{i,t}$) with the resolution of $1/\Gamma$ and $1^\circ$C, respectively.

Let $M_{\mathrm{tr}}:=|\mathcal M^{\mathrm{train}}|$, $M_{\mathrm{inf}}:=|\mathcal M^{\mathrm{inf}}|$, and $H:=|\Delta_m|$. For $N$ AIDCs, the action-space dimension of the WM agent for training and inference jobs is expressed as
\begin{equation}
|\mathcal A^{\mathrm{WM}}(N)|
=
\left[\binom{N+\Gamma-1}{\Gamma}\right]^{M_{\mathrm{tr}}+M_{\mathrm{inf}}}.
\label{eq:wm_complexity}
\end{equation}

For each AIDC, the temporal-shifting action $\boldsymbol q_{i,t}$ includes $H$ deferral options for each class-$m$ training job, and the cardinality of such action space is written as  
\begin{equation}
|\mathcal A^{q}|
=
\left[\binom{H+\Gamma-1}{\Gamma}\right]^{M_{\mathrm{tr}}}
.
\label{eq:q_complexity}
\end{equation}

The GPU allocation is conducted using the actions $(\boldsymbol u_{i,t},\boldsymbol v_{i,t})$ under $\sum_{m\in\mathcal M^{\mathrm{train}}}u_{m,i,t} + \sum_{m\in\mathcal M^{\mathrm{inf}}}v_{m,i,t} \le 1$. Introducing a slack variable converts this inequality into a simplex with $M_{\mathrm{tr}}+M_{\mathrm{inf}}+1$ components. Hence, the cardinality of the action space for GPU allocation is expressed as
\begin{equation}
|\mathcal A^{g}|
=
\binom{M_{\mathrm{tr}}+M_{\mathrm{inf}}+\Gamma}{\Gamma}.
\label{eq:g_complexity}
\end{equation}

Along with the action-space cardinality $|\mathcal{A}^{T}|$ of the CRAC supply temperature $T^{\mathrm{CRAC,sup}}_{i,t}$ with a predefined resolution of $1^\circ$C, the action-space cardinality of each AIDC is calculated as
\begin{equation}
|\mathcal A^{\mathrm{AIDC}}|
=
|\mathcal A^{q}|\,|\mathcal A^{g}|\,|\mathcal A^{T}|.
\label{eq:aidc_complexity}
\end{equation}

Finally, based on \eqref{eq:wm_complexity} and \eqref{eq:aidc_complexity}, the cardinality $|\mathcal A_{\mathrm{flat}}(N)|$ of the joint action space in the flat architecture is expressed as the Cartesian product of the WM actions and all local AIDC actions as follows:
\begin{equation}
|\mathcal A_{\mathrm{flat}}(N)|
=
|\mathcal A^{\mathrm{WM}}(N)|\cdot |\mathcal A^{\mathrm{AIDC}}|^N.
\label{eq:flat_complexity}
\end{equation}

By contrast, the proposed hierarchical method factorizes the entire policy into one WM policy and $N$ conditioned local AIDC policies. Since the MAT architecture generates actions autoregressively in a sequential decoding process, the neural network parameterizes these local action spaces additively at each decoding step, rather than exploring their full Cartesian product simultaneously. This results in a significant reduction in the effective policy search complexity. Accordingly, the cardinality $|\mathcal A_{\mathrm{hier}}(N)|$ of the action space in the hierarchical architecture is expressed as 
\begin{equation}
|\mathcal A_{\mathrm{hier}}(N)|
=
|\mathcal A^{\mathrm{WM}}(N)|+N|\mathcal A^{\mathrm{AIDC}}|.
\label{eq:hier_complexity}
\end{equation}

Note that for a fixed $\Gamma$ the cardinality of action space for the flat structure grows as $\mathcal{O}\!\left(N^{(M_{\mathrm{tr}}+M_{\mathrm{inf}})\Gamma}\,|\mathcal A^{\mathrm{AIDC}}|^N\right)$, whereas the effective complexity of the hierarchical structure grows as $\mathcal{O}\!\left(N^{(M_{\mathrm{tr}}+M_{\mathrm{inf}})\Gamma}+N|\mathcal A^{\mathrm{AIDC}}|\right)$. Therefore, the proposed hierarchical CA-MARL framework transforms an exponentially growing joint action space into a tractable and factorized decision process, thereby enabling scalable learning in a large-scale DSO--AIDC operator coordination.

\begin{table}[t!]
\centering
\caption{Simulation parameters}
\label{tab:all_params}
\renewcommand{\arraystretch}{1.2}
\footnotesize
\begin{tabular}{l|l}
\noalign{\hrule height 1.2pt}
\multicolumn{2}{c}{\textbf{CEF-Integrated Power Distribution System Model}}\\
\noalign{\hrule height 0.8pt}
\textbf{Symbol} & \textbf{Value} \\
\hline
$(\underline{V}, \overline{V})$ & $(0.9, 1.1)$ p.u. \\
$P_i^{\mathrm{gt}}$ & $[0, 500]$ kW (Nodes 6, 25, 30) \\
$Q_i^{\mathrm{gt}}$ & $[-300, 300]$ kVAr (Nodes 6, 25, 30) \\
$w_i^{\mathrm{gt}}$ & $[0.30, 0.35, 0.65, 0.90]$ kgCO$_2$/kWh (Nodes 0, 6, 25, 30) \\
$\kappa_t$ &
\begin{tabular}[t]{@{}l@{}}
Base: 0.1 \$/kWh, Peak [11 a.m., 3 p.m.]: 0.3 \$/kWh,\\
Off-peak [7 p.m., 11 p.m.]: 0.03 \$/kWh
\end{tabular} \\
$\lambda$ & $0.01$ \\
$\widehat{P}_{i,t}^{\mathrm{load}}$ & IEEE 33 load $\times$ normalized profile \\
$\eta$ & $0.2$ (reactive-to-real power ratio) \\
$\Delta t$ & $15$ min (DSO problem), $1$ min (MARL control) \\
\noalign{\hrule height 1.2pt}
\multicolumn{2}{c}{\textbf{AIDC Model}}\\
\noalign{\hrule height 0.8pt}
\textbf{Symbol} & \textbf{Value} \\
\hline
$\mathcal{N}^{\text{AIDC}}$ & 3 AIDCs (Nodes 8, 28, 32) \\
$G_i$ & $400$ GPUs (50 racks $\times$ 8 GPUs) \\
$P_i^{\mathrm{idle}}$ & $100$ kW \\
$\alpha^{\mathrm{train}}, \alpha^{\mathrm{inf}}$ & ${7.0 \times 10^{-4}}$ (train), ${1.68 \times 10^{-4}}$ (inference)  kW/TOPS \\
$\xi_{\max}^{\mathrm{train}}$ & ${6.0\times10^4}$ TO/GPU/min \\
$\xi_{\max}^{\mathrm{inf}}$ & ${2.5\times10^5}$ TO/GPU/min \\
$j^{\mathrm{train}}_{m}$ & $6.91\times10^8$ (LLM), $2.88\times10^8$ (VAE) TO/job \\
$j^{\mathrm{inf}}_{m}$ & $3.01\times10^7$ (DeepResearch), $1.5\times10^7$ (Search) TO/job \\
$b_m$ & $16$ (LLM), $8$ (VAE) GPUs \\
$\tau^{\mathrm{train}}_m$ & $720$ min \\
$\Delta_m$ & $(0, 120, 240, 360, 480, 600)$ min \\
$\tau^{\mathrm{inf}}_m$ & $30$ min (DeepResearch), $15$ min (Search) \\
$\bar{\lambda}^{\mathrm{train}}_m$ & $0.0585$ (LLM), $0.0840$ (VAE) jobs/min \\
$A^{\mathrm{train}}_m$ & $0.0117$ (LLM), $0.0210$ (VAE) jobs/min \\
$\Theta^{\mathrm{train}}_m$ & $120$ (LLM), $90$ (VAE) min \\
$\phi^{\mathrm{train}}_m$ & $0.0$ (LLM, VAE) \\
$\sigma^{\mathrm{train}}_{\epsilon}$ & $0.0062$ (LLM), $0.0168$ (VAE) jobs/min \\
$\bar{\lambda}^{\mathrm{inf}}_m$ & $16.5$ (DeepResearch), $110.0$ (Search) jobs/min \\
$A^{\mathrm{inf}}_m$ & $1.65$ (DeepResearch), $8.25$ (Search) jobs/min \\
$\Theta^{\mathrm{inf}}_m$ & $60$ (DeepResearch), $30$ (Search) min \\
$\phi^{\mathrm{inf}}_m$ & $10.0$ (DeepResearch), $5.0$ (Search) \\
$\sigma^{\mathrm{inf}}_{\epsilon}$ & $0.825$ (DeepResearch), $2.75$ (Search) jobs/min \\

\noalign{\hrule height 1.2pt}
\multicolumn{2}{c}{\textbf{CRAC-Based Cooling Model}}\\
\noalign{\hrule height 0.8pt}
\textbf{Symbol} & \textbf{Value} \\
\hline
$\Delta T^{\mathrm{sup}}_{i,r}$ & $4.0^\circ$C \\
$\Delta T^{\mathrm{CRAC,ret}}_{i,r}$ & $3.7^\circ$C \\
$T^{\mathrm{CRAC,sup}}_{i}$ & $[18, 23]^\circ$C \\
$\Delta T^{\mathrm{CT,air}}_{i,t}$ & $1.0^\circ$C \\
$\overline{T}^{\mathrm{inlet}}_{i,r}$ & $27^\circ$C \\
$c^{\mathrm{air}}$ & $1006$ J/(kg$\cdot^\circ$C)\\
$\rho^{\mathrm{air}}$ & $1.225$ kg/m$^3$ \\
$\Phi^{\mathrm{sfan}}$ & $6.0$ m$^3$/s \\
$\operatorname{COP}_i(T)$ & $0.0068T^2+0.008T+0.458$ \\
$P^{\mathrm{CT,REF}}_i$ & $6.0$ kW \\
$\Phi^{\mathrm{CT,REF}}_i$ & $8.5$ m$^3$/s \\
\noalign{\hrule height 1.2pt}
\end{tabular}

\end{table}

\begin{table*}[ht!]
\centering
\caption{Performance of MAPPO, Transformer, MAT-Dec, and proposed MAT methods in both joint and power modes.}
\label{tab:results}
\renewcommand{\arraystretch}{1.15}
\setlength{\tabcolsep}{4pt}
\begin{tabular}{l|cc|cc|cc|cc}
\noalign{\hrule height 1.2pt}
\multirow{2}{*}{\textbf{Metric}}
 & \multicolumn{2}{c|}{\textbf{MAPPO}}
 & \multicolumn{2}{c|}{\textbf{Transformer}}
 & \multicolumn{2}{c|}{\textbf{MAT-Dec}}
 & \multicolumn{2}{c}{\textbf{MAT (proposed)}}  \\
 & \textbf{Joint} & \textbf{Power}
 & \textbf{Joint} & \textbf{Power}
 & \textbf{Joint} & \textbf{Power}
 & \textbf{Joint} & \textbf{Power} \\
\noalign{\hrule height 1.2pt}

Avg. Total Training/Inference Throughput [TOPS] ($\texttt{R}_{1,t}$)
 & 59.86M & 63.73M
 & 54.24M & 55.62M
 & 64.60M & 69.65M
 & 65.67M & \textbf{70.69M} \\

Total Electricity Purchase Cost [\$] ($\texttt{R}_{2,t}$)
 & 1886.47 & 1960.69
 & \textbf{1859.21} & 1866.52
 & 2142.61 & 2249.08
 & 2084.77 & 2554.95 \\

Total Carbon Emissions [tonCO$_2$] ($\texttt{R}_{3,t}$)
 & 10.54 & 10.80
 & 10.58 & 10.67
 & 10.01 & 10.94
 & \textbf{9.77} & 12.09 \\

Total Number of Dropped Training+Inference Jobs ($\texttt{R}_{4,t}+\texttt{R}_{5,t}$)
 & 10627 & 6196
 & 14885 & 7617
 & 6207 & 401
 & 419 & \textbf{45} \\

Avg. Carbon Emission Efficiency [MWh/tonCO$_2$]
 & 1.65 & 1.60
 & 1.68 & 1.62
 & 1.67 & 1.62
 & \textbf{1.69} & 1.67 \\

Avg. PUE
 & 1.16 & 1.15
 & 1.15 & 1.14
 & 1.13 & 1.18
 & \textbf{1.08} & 1.39 \\
 
\noalign{\hrule height 1.2pt}
\end{tabular}
\end{table*}

\begin{figure}[t!]
    \centering
    \includegraphics[width=0.9\linewidth]{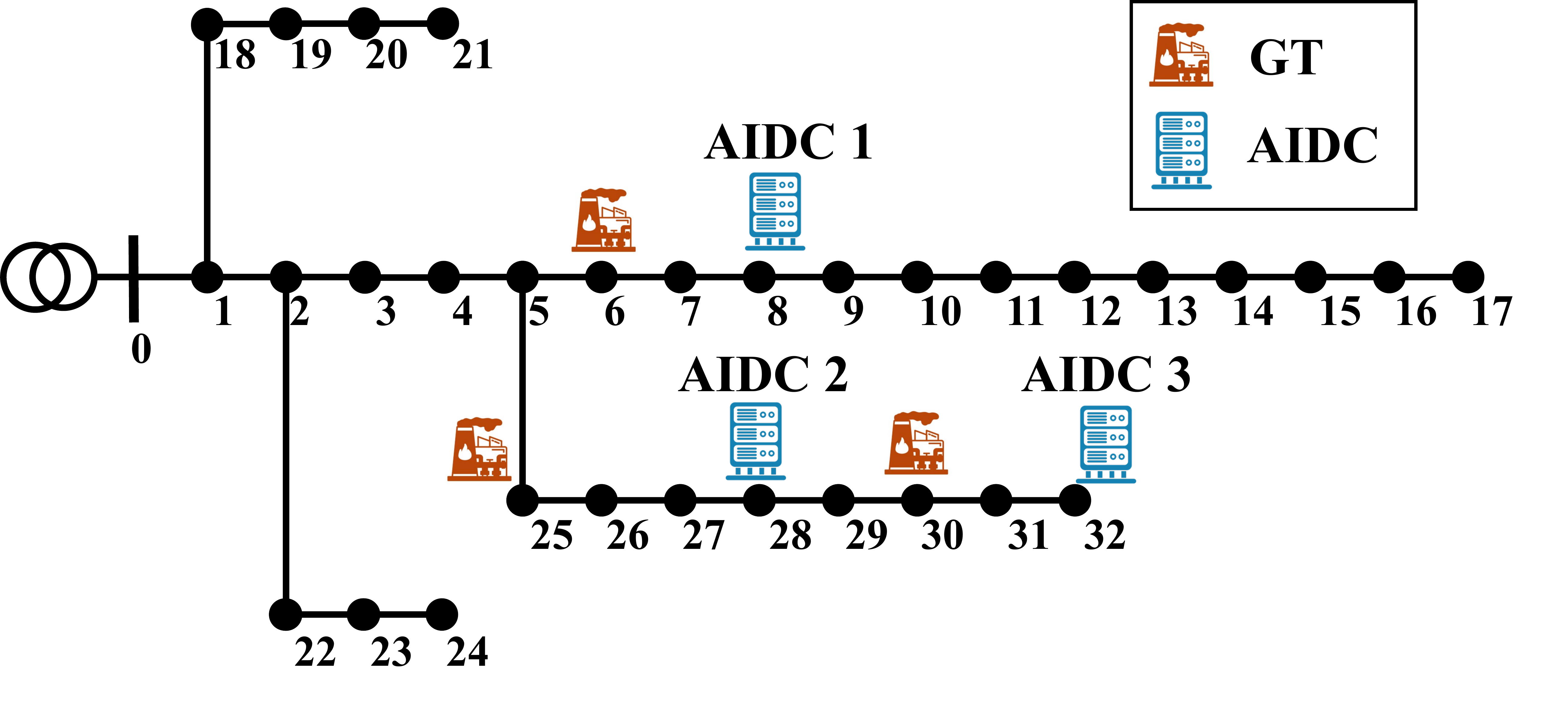}
    \caption{IEEE 33-node power distribution system.}
    \label{fig:33bus}
\end{figure}

\section{Performance Evaluation} \label{sec:performance}
\subsection{Experimental Setup}\label{subsec:exp_setup}
The performance of the proposed CA-MARL framework based on the MAT method was assessed using an IEEE 33-node power distribution system including three GTs and three AIDCs as shown in Fig.~\ref{fig:33bus} \cite{feeders}. 
Each AIDC represents a regional-scale facility connected to the IEEE 33-node power distribution system with a nominal voltage of 12.66 kV. The modeled AIDCs have an average power consumption on the order of several megawatts (approximately 8–12 MW), corresponding to medium-scale AI computing facilities with several hundred GPUs.. Table~\ref{tab:all_params} lists the parameter values used
in the proposed framework. To ensure practical relevance, the simulation parameters listed in Table~\ref{tab:all_params} were configured based on empirical studies and then scaled to align with the capacity of the IEEE 33-node power distribution system.
Specifically, the computational requirements and energy consumption of the AI workloads were modeled based on neural scaling laws~\cite{Kaplan2020ScalingLaws}.
Four types of AIDC workloads were considered: i) LLM and variational autoencoder (VAE) for delay-tolerant training jobs, and ii) DeepResearch and Search for delay-sensitive inference jobs.
The total number of arriving AIDC workloads $N^k_{m,t}$ in type $k$ and class $m$ was generated by the following time-varying stochastic arrival rate $\lambda^k_m(t)$ calibrated to AIDC workload traffic patterns:
\begin{equation}
\lambda^k_m(t) = \bar{\lambda}^k_m + A^k_m \sin\!\left(\frac{2\pi t}{\Theta^k_m} + \phi^k_m\right) + \epsilon_t,
\end{equation}
where for class $m$-based workload $\bar{\lambda}_m ^k$ is the average arrival rate, $A_m ^k$, $\Theta_m ^k$, and $\phi_m ^k$ are the amplitude, period, and phase shift of the sinusoidal function, and $\epsilon_t$ is an additive noise based on the normal distribution with zero mean.
The workload generation model reflects the stochastic and non-stationary characteristics of AIDC workloads through time-varying demand patterns with random fluctuations. In particular, bursty training workloads derive from temporal concentration and accumulation of workload arrivals, while inference workloads exhibit distributed and high-frequency variations due to phase-shifted demand patterns.
As listed in Table~\ref{tab:all_params}, training jobs (LLM, VAE) exhibit slow time-varying yet bursty patterns, whereas inference jobs (DeepResearch, Search) exhibit high-frequency fluctuations along with a broad range of workload traffic via interactive user queries. The CRAC-based cooling system operates under standard thermal guidelines~\cite{sustaindc}; a quadratic COP model is employed to capture the trade-off between chiller efficiency and cooling capacity. The agents schedule AIDCs at a 1-min resolution, whereas the DSO computes the NCI states at a 15-min resolution; thus, each NCI value is held constant across the 1-min steps within the corresponding 15-min interval. This multi-time-scale architecture allows the agents to respond to NCI signals at a finer resolution while maintaining computational tractability at the DSO level, and ensures stable interaction between the DSO and AIDC agents under piecewise-constant NCI signals.

We demonstrate the advantages of the proposed MAT-based method by comparing the following three baseline methods: i) MAT-Dec method with decentralized decoders while retaining the shared encoder, ii) Transformer (parallel attention) method without a shared encoder, and iii) multi-agent PPO (MAPPO) method as an MARL benchmark using multi-layer perceptron (MLP)-based policies. These baselines were selected to reflect the two key challenges of the proposed framework: i) temporal dependency modeling and ii) multi-agent spatial coordination. Transformer effectively models temporal dependencies but lacks explicit spatial coordination among AIDCs, while MAPPO enables multi-agent coordination but is limited in capturing long-term temporal workload shifting. Therefore, these baselines  represent complementary limitations, and their comparison demonstrates that both capabilities are required simultaneously. The superior performance of MAT arises from its ability to jointly capture temporal dependencies and coordinated multi-agent decision-making  in the proposed framework.
In addition, to investigate the impact of carbon-aware functionality in the proposed framework on low-carbon AIDC operation, the performance of a joint mode (i.e., the proposed method considering both economical and eco-friendly AIDC operations) is compared with a power mode considering only the economic operations of AIDCs, excluding carbon emission reduction. The power mode excludes the contribution of the NCI signal used in the joint mode by deleting the total carbon emissions $J_2$ of GTs and substation in~\eqref{eq:obj} ($\lambda=0$), CEF constraint~\eqref{eq:cefbalance}, NCI state in ~\eqref{eq:state_wm_agent},~\eqref{eq:state_aidc_agent}, and total carbon emissions $\texttt{R}_{3,t}$ of AIDCs ($c_3=0$).

\begin{figure}[t!]
    \centering
    \includegraphics[width=0.95\linewidth]{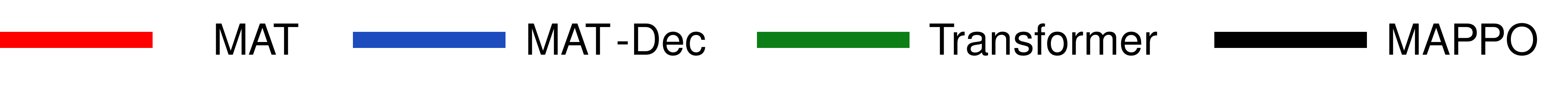} 
        \centering
        \includegraphics[width=0.98\linewidth]{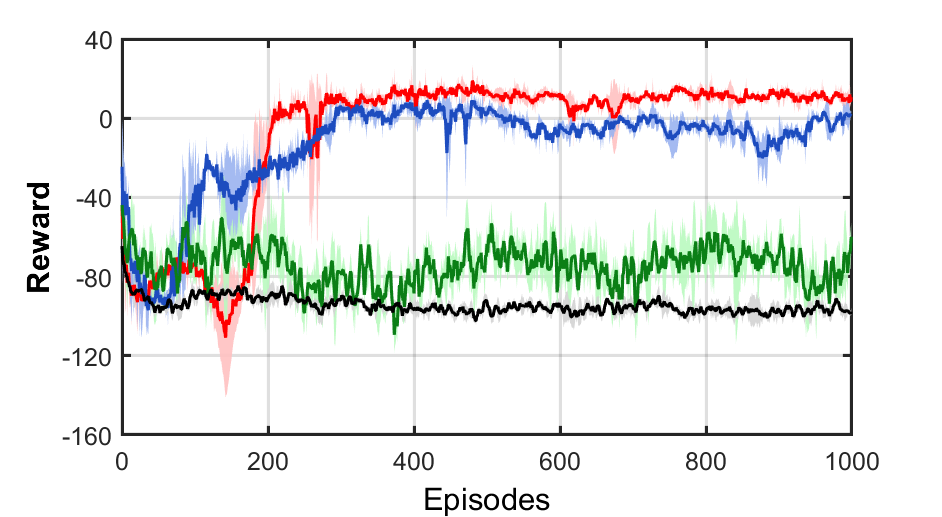}
        \caption{Comparison of rewards among the MAPPO, Transformer, MAT-Dec, and proposed MAT methods in a joint mode.}
        \label{fig:reward_joint}
\end{figure}

\subsection{Evaluation Results}\label{subsec:results}
Fig.~\ref{fig:reward_joint} compares the training curves that exhibit convergence of the rewards between the MAPPO, Transformer, MAT-Dec, and proposed MAT methods in joint mode. Training is performed over 1000 episodes, where each solid line represents the mean reward over three independent random seeds and the shaded regions indicate the corresponding ±1\% standard deviation across runs, thereby illustrating the convergence behavior and variability of each method.
Note that the two MAT-based methods (MAT and MAT-Dec) achieve significantly higher rewards than the Transformer and MAPPO methods. This superiority stems from the autoregressive policy generation of the MAT framework, which enables sequential and conditioned decision-making to minimize conflicts. By contrast, the Transformer and MAPPO methods generate simultaneous actions, leading to conflicting decisions. In addition, the MAT-Dec method exhibits a lower performance than the proposed MAT. This is because the absence of joint decoding in the MAT-Dec method causes the training data to be fragmented across decentralized networks, thereby reducing training efficiency.
These results demonstrate that the proposed MAT method effectively allows agents to implicitly coordinate their spatial and temporal workload shifting strategies, resolving inter-agent conflicts and maximizing system-wide performance.

Table~\ref{tab:results} compares the performance of three baseline and proposed methods over a 24-hour period (1,440 steps) in terms of average total throughput ($\texttt{R}_{1,t}$), total electricity purchase cost ($\texttt{R}_{2,t}$), total carbon emissions ($\texttt{R}_{3,t}$), total number of dropped training/inference jobs ($\texttt{R}_{4,t}+\texttt{R}_{5,t}$), average carbon emission efficiency, and average power usage effectiveness (PUE).
Carbon emission efficiency is the ratio of total AIDC electricity consumption to carbon emissions, and PUE is defined as $P^{\mathrm{AIDC}}_{i,t} / P^{\mathrm{IT}}_{i,t}$. For the former metric, a higher value indicates greater carbon reduction efficiency. By contrast, for the latter, a value closer to one indicates a higher cooling efficiency. 
Several observations can be made from Table~\ref{tab:results}.
First, compared to the power mode, which maximizes job throughput and minimizes the number of dropped jobs, the joint mode for all four methods achieves a more eco-friendly AIDC operation, resulting in a reduction in total carbon emissions and an enhancement of carbon emission efficiency. This demonstrates that the NCI signal, along with its related environment and reward function in the joint mode, successfully contributes to the low-carbon operation of AIDCs. In addition, the power mode yields a higher total electricity purchase cost than the joint mode because it requires more electricity from the grid to obtain a higher job throughput by utilizing more GPU resources.
Second, for the joint mode, the proposed MAT method exhibits lower total carbon emissions and higher carbon emission efficiency and throughput, with a significantly reduced job drop compared to the MAPPO, Transformer, and MAT-Dec methods. This confirms that the MAT method coordinates spatio-temporal decisions more effectively than the other methods to resolve inter-agent conflicts. 
Third, the proposed MAT method in the joint mode exhibits the lowest PUE value. This demonstrates that the proposed method successfully improves cooling efficiency by managing the trade-off between IT load distribution and cooling power consumption while ensuring low-carbon operation of AIDC.

\begin{figure}[t!]
    \centering
        \centering
        \includegraphics[width=0.96\linewidth]{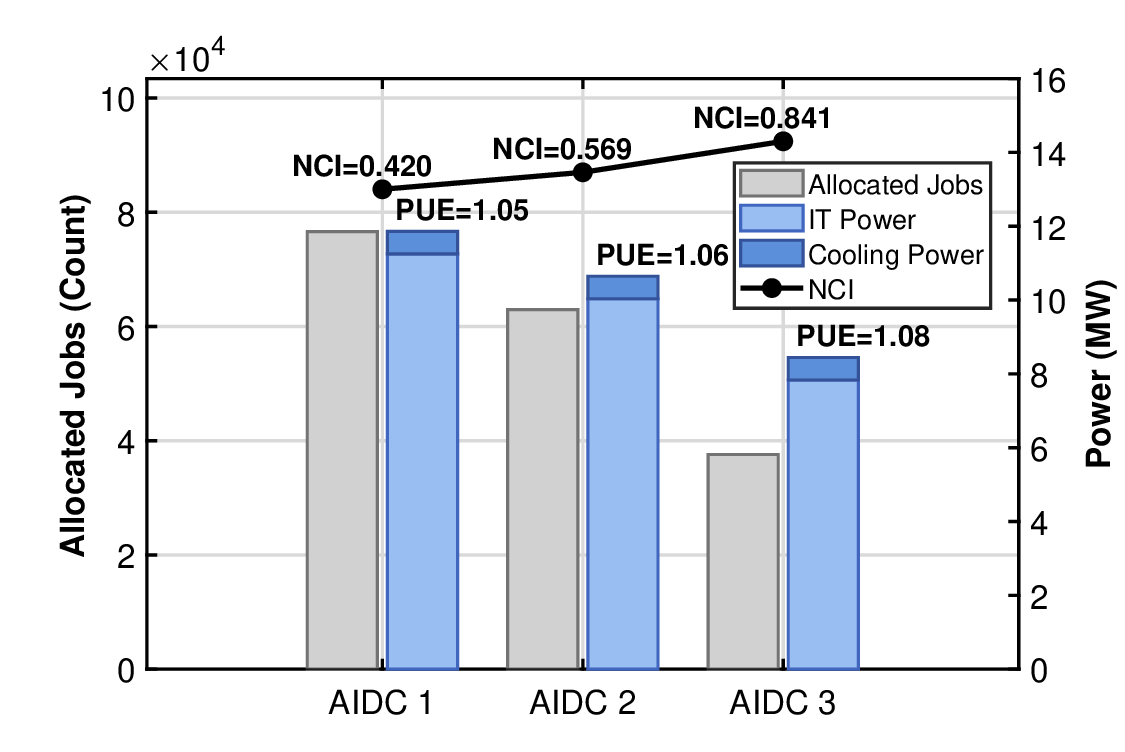}
        \caption{Allocated training/inference jobs and power consumption at three AIDCs under varying NCIs in a joint mode of the proposed MAT method.}
        \label{fig:allocation}
\end{figure}

\begin{figure}[t!]
    \centering
    \includegraphics[width=0.7\linewidth]{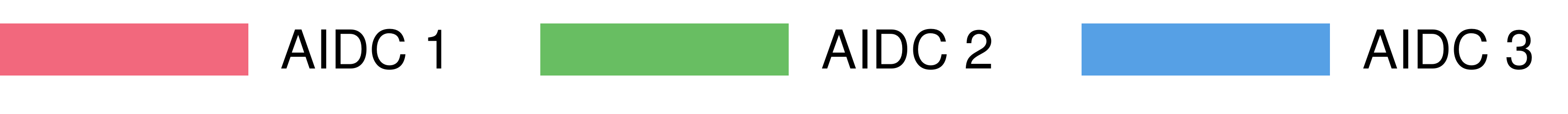}
    \subfloat[NCI variations.]{
        \includegraphics[width=0.95\linewidth]{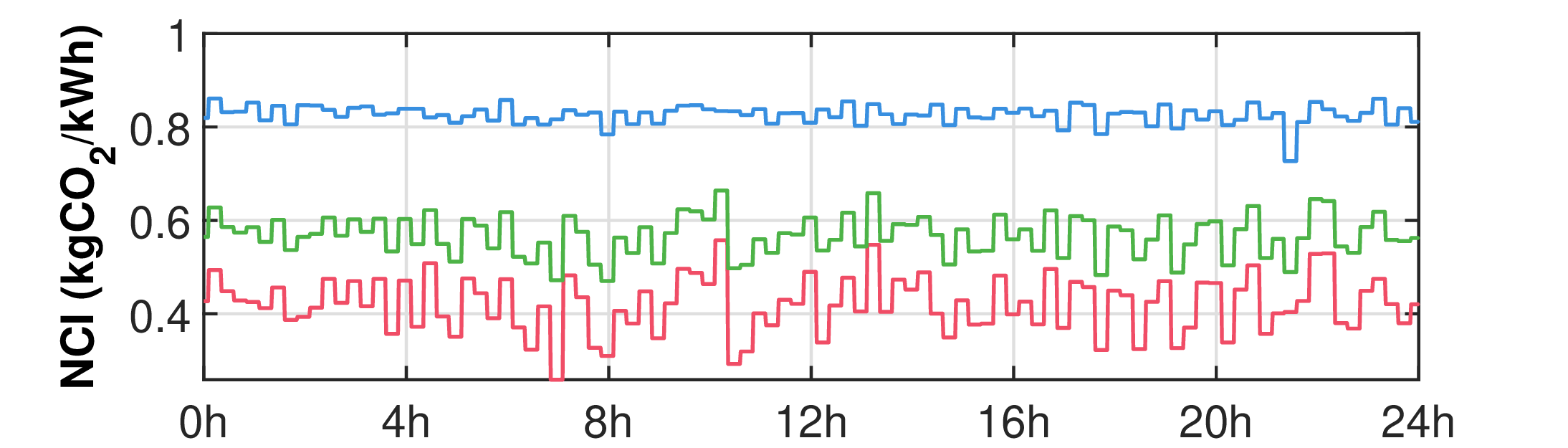}
        \label{fig:nci}
    }\\
    \subfloat[Spatial training/inference job shifts by the WM agent.]{
        \includegraphics[width=0.95\linewidth]{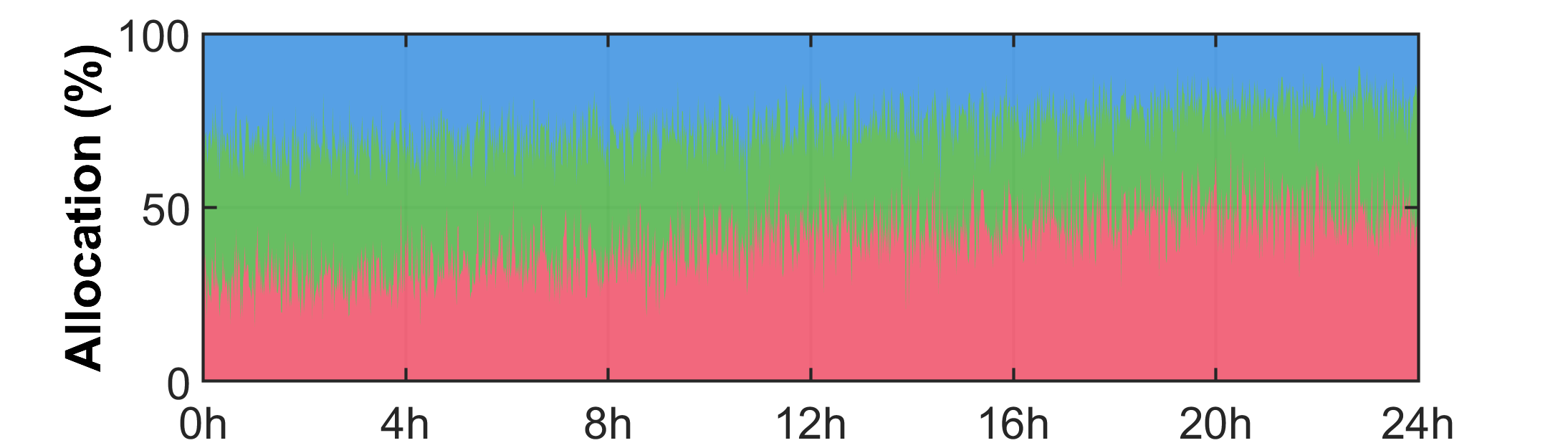}
        \label{fig:ss}
    }\\
    \subfloat[Temporal training job shifts by the AIDC agents.]{
        \includegraphics[width=0.95\linewidth]{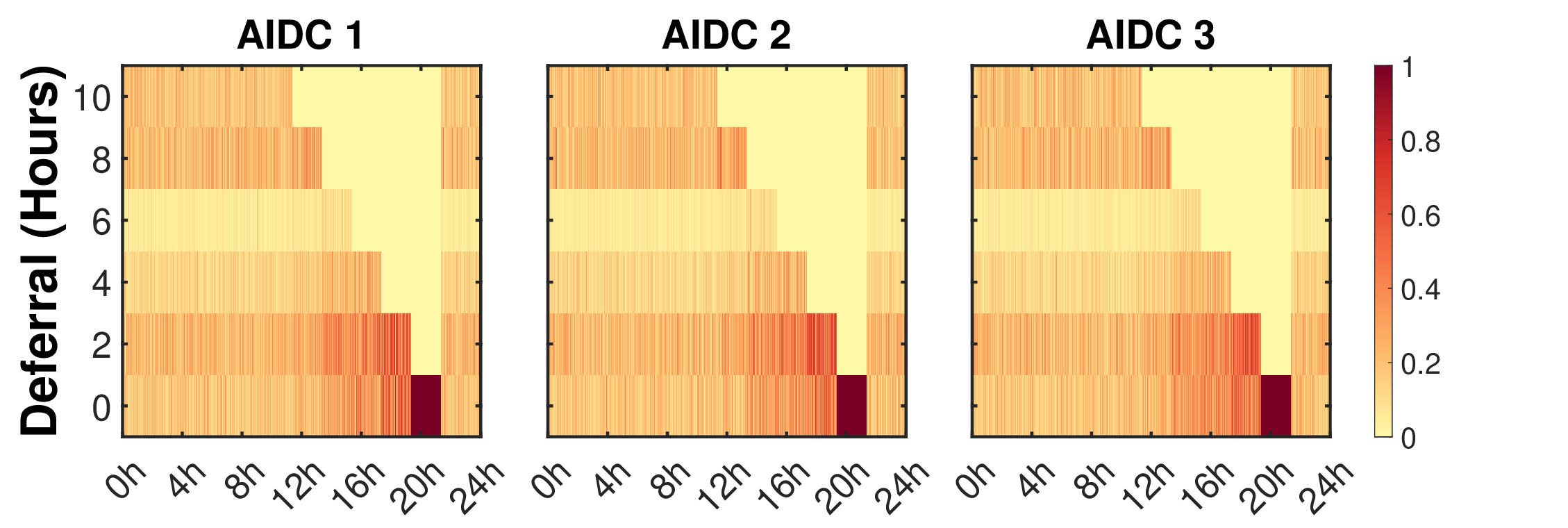}
        \label{fig:ts}
    }
    \caption{Spatio-temporal training/inference job shifting for three AIDCs under varying NCIs during a day.}
    \label{fig:nci_spatial_temporal}
\end{figure}

Fig.~\ref{fig:allocation} compares the training/inference jobs allocated by the WM agent and the power consumption (IT+cooling power consumption) of the three AIDCs with varying NCIs in the joint mode of the proposed MAT method. 
Note that an increase in the NCI leads to a decrease in the number of allocated jobs and the power consumption of AIDCs. Among the three AIDCs, AIDC 1 with the lowest NCI of 0.420 was assigned the largest number of jobs, whereas AIDC 3 with the highest NCI of 0.841 was assigned the lowest. In addition, AIDCs 1 and 3 achieved the best and worst performances, respectively, in terms of PUE. These observations demonstrate that the WM agent in the proposed method maintains the low-carbon operation of AIDCs by adaptively responding to variations in NCI while ensuring operation in terms of power consumption. 

Fig.~\ref{fig:nci_spatial_temporal} shows the spatio-temporal training/inference job-shifting capability of the WM and AIDC agents during a day when the three AIDCs have different NCI profiles.
Considering the heterogeneous NCI profiles depicted in Fig.~\ref{fig:nci}, Fig.~\ref{fig:ss} shows that the allocation ratios of training jobs for AIDCs 1 and 3 increase and decrease, respectively, whereas the allocation ratio of training jobs for AIDC 2 does not change significantly. This is because the NCIs of AIDC 1 are much lower than those of AIDC 3; therefore, more jobs are assigned to AIDC 1 to reduce carbon emissions.
Fig.~\ref{fig:ts} illustrates the temporal job-shifting capabilities of the three AIDC agents using heatmaps. Note that the arriving jobs of AIDCs during the peak period of TOU prices [11 a.m., 3 p.m.] are deferred to an off-peak period of TOU prices [7 p.m., 11 p.m.]. This observation confirms that the AIDC agents successfully contribute to the economic operation of AIDCs. Furthermore, since the AIDC agents are trained under the same policy with identical TOU pricing signals, they consistently learn similar temporal workload shifting behaviors. Consequently, their distributions show similar patterns across AIDCs. Nevertheless, slight statistical differences are observed due to dynamics and stochastic variations of AIDCs, as reflected in the mean and variance of the deferral times (3.6231 h and 3.3307 for AIDC 1; 3.6090 h and 3.3300 for AIDC 2; 3.6442 h and 3.4342 for AIDC 3).

\begin{figure}[t!]
    \centering
    \includegraphics[width=0.97\linewidth]{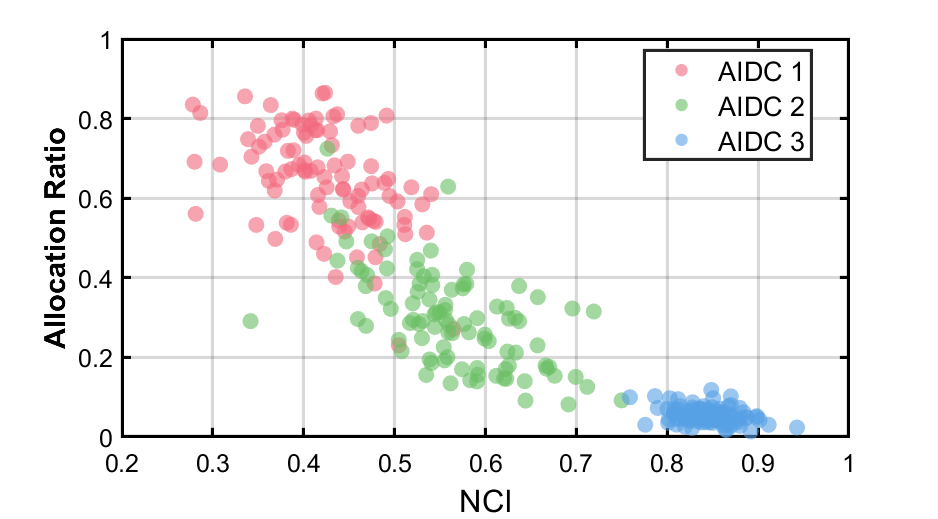}
    \caption{Relationship between NCI and workload allocation ratio for each AIDC in the joint mode.}
    \label{fig:scatter}
\end{figure}

\begin{figure}[t!]
    \centering
    \includegraphics[width=0.7\linewidth]{Figures/aidclegends.pdf}\\
    \includegraphics[width=0.97\linewidth]{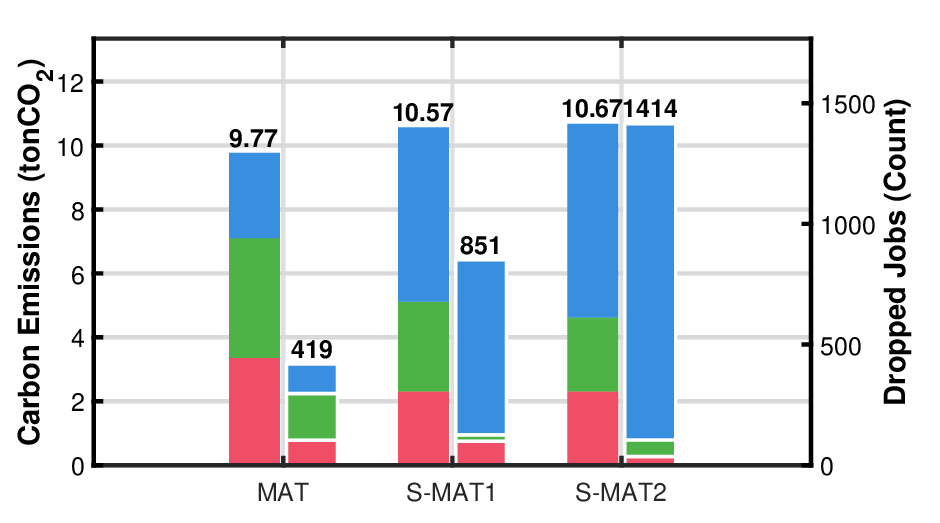}    
    \caption{Performance comparison of the MAT method between static and dynamic job shifting of the WM agent.}
    \label{fig:static}
\end{figure}

\begin{figure}[t!]
    \centering
    \begin{subfigure}{1.0\linewidth}
        \centering
        \includegraphics[width=0.99\linewidth]{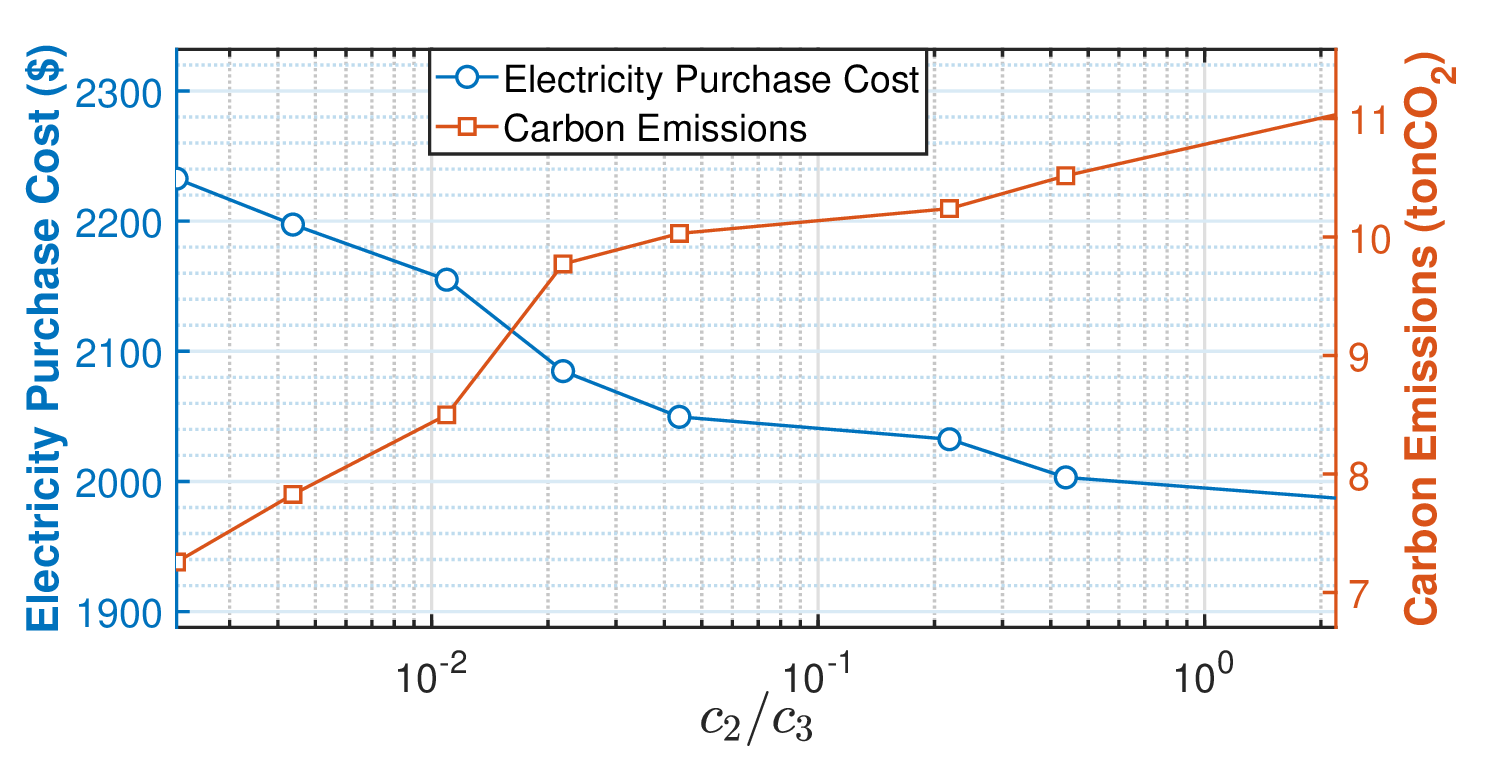}
        \caption{Sensitivity to the weight ratio $c_2/c_3$ in the reward function.}
        \label{fig:sens_c2c3}
    \end{subfigure}
    \begin{subfigure}{0.99\linewidth}
        \centering
        \includegraphics[width=0.99\linewidth]{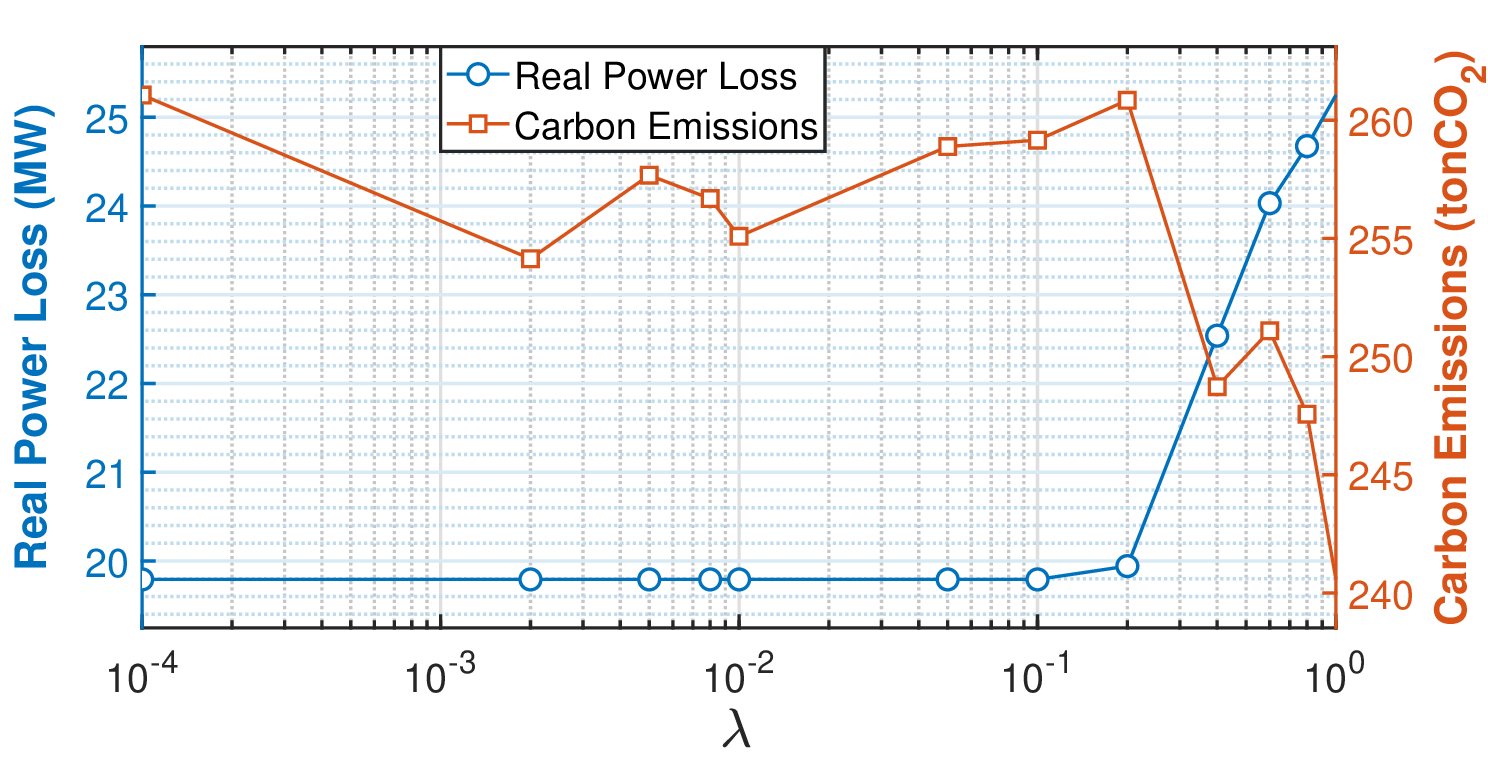}
        \caption{Sensitivity to the penalty parameter $\lambda$ in the objective function.}
        \label{fig:sens_lambda}
    \end{subfigure}
    \caption{Sensitivity analysis of the reward and objective functions for the AIDC operator and DSO with respect to changes in the weight ratio and penalty parameters.}
    \label{fig:sensitivity}
\end{figure}

\begin{figure}[t!]
    \centering
    \includegraphics[width=0.7\linewidth]{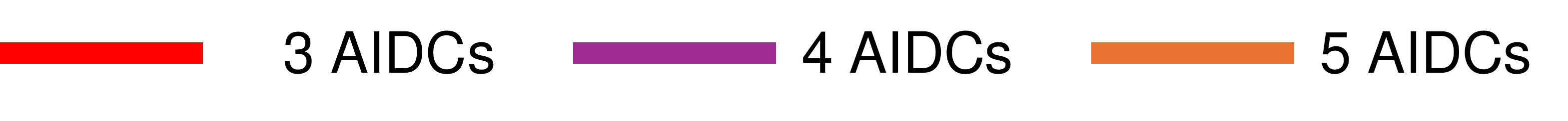} 
        \centering
        \includegraphics[width=0.98\linewidth]{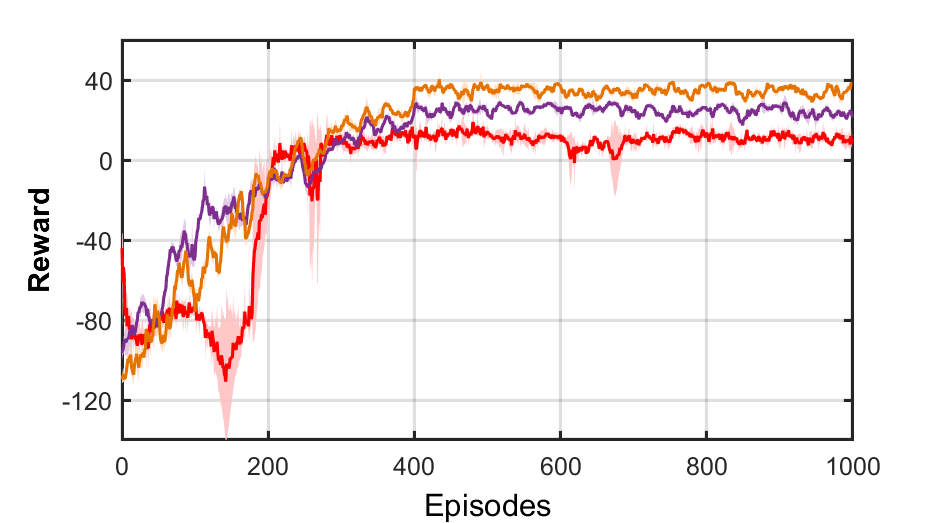}
        \caption{Comparison of rewards for the proposed MAT method with different numbers of AIDCs (3, 4, and 5) in the joint mode.}
        \label{fig:reward_convergence_345}
\end{figure}

To explicitly demonstrate the spatial effect of the CEF-induced NCI signal, Fig.~\ref{fig:scatter} illustrates the relationship between the NCI and the workload allocation ratio for each AIDC in the joint mode. Here, the workload allocation ratio represents the proportion of arriving workloads assigned to each AIDC by the WM agent. As shown in Fig.~\ref{fig:scatter}, an overall inverse relationship is observed, where AIDCs with lower NCI values receive higher workload allocations, while those with higher NCI values receive fewer workloads. This indicates that the WM agent adaptively allocates workloads in response to CEF-induced NCI variations, thereby achieving spatially differentiated carbon-aware scheduling.

Fig.~\ref{fig:static} highlights that the proposed MAT method with dynamic spatial shifting of jobs outperforms the method with static spatial shifting of jobs in terms of reduction in carbon emissions and dropped jobs. In this figure, S-MAT1 and S-MAT2 represent the static MAT methods in which the WM agent allocates the jobs to AIDCs 1, 2, and 3 according to fixed ratios of 1:2:7 and 1:1:8, respectively, without considering the operating conditions of CEF-integrated power distribution grids. Note that the proposed MAT method calculates lower values of carbon emissions and fewer dropped jobs than the S-MAT1 and S-MAT2 methods, as indicated in this figure.
Furthermore, as the fixed allocation ratio becomes more imbalanced, carbon emissions increase, and the number of dropped jobs increases. These observations demonstrate that the WM agent in the proposed MAT method enables the low-carbon and high-throughput operation of AIDCs through the dynamic shifting of jobs.

Fig.~\ref{fig:sensitivity}(a) shows the results of the total electricity purchase cost ($\texttt{R}_{2,t}$) and total carbon emissions ($\texttt{R}_{3,t}$) over the entire scheduling period with respect to weight ratio $c_2/c_3$ in the reward function of the AIDC operators. As shown in Fig.~\ref{fig:sensitivity}(a), we verify a trade-off relationship between economic and environment performances. Specifically, a higher (or lower) $c_2/c_3$ leads to a lower (or higher) total electricity purchase cost; however, it yields a higher (or lower) total carbon emission. In this study, $c_2/c_3=0.02$ is selected as a practically balanced setting, which avoids both an excessively cost-driven policy and an overly carbon-dominant policy.

Fig.~\ref{fig:sensitivity}(b) shows the sensitivity of total real power loss ($J_1$) and total carbon emissions ($J_2$) with respect to $\lambda$ in the objective function~\eqref{eq:obj}. Note from this figure that the effect of $\lambda$ is region-dependent, rather than exhibiting a uniform trade-off over the entire range. In the low-$\lambda$ region ($\lambda\leq 10^{-1}$), the objective function is heavily dominated by real power loss minimization, yielding nearly constant $J_1$ and limited sensitivity of $J_2$, where fluctuations arise from multiple near-optimal dispatch solutions of GTs. By contrast, when $\lambda > 10^{-1}$, the carbon emission term starts to influence dispatch solutions, leading to reduced $J_2$ at the expense of increased $J_1$. This demonstrates that a clear economic-environmental trade-off emerges only beyond a certain threshold of $\lambda$. In this simulation, $\lambda=0.01$ is selected as a conservative operating point, where carbon-awareness is incorporated without significantly degrading system efficiency.

\begin{figure}[t!]
    \centering
        \centering
        \includegraphics[width=0.98\linewidth]{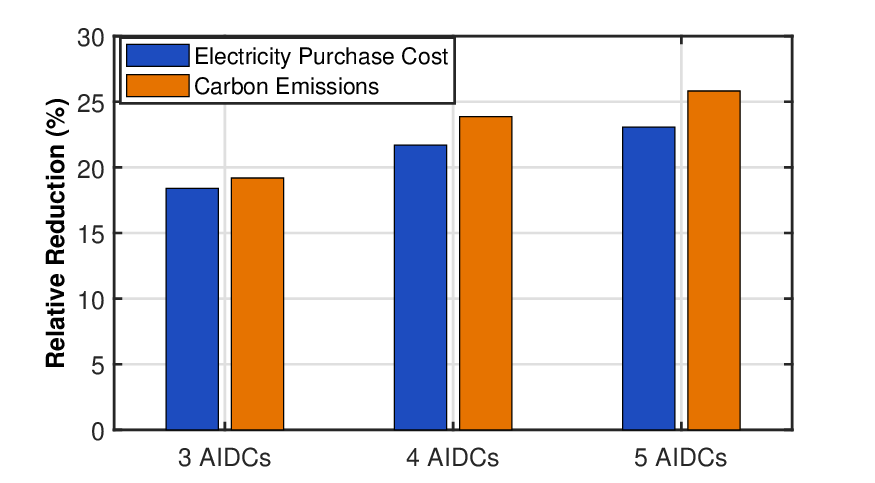}
        \caption{Relative reduction of electricity purchase cost and carbon emissions in the proposed MAT method using the joint mode with respect to the power mode with different numbers of AIDCs.}
        \label{fig:reward_345}
\end{figure}

Fig.~\ref{fig:reward_convergence_345} compares the training curves over 1000 episodes for the proposed MAT method with different numbers of AIDCs in the joint mode. For the cases with four and five AIDCs, additional AIDCs are connected to node 20 and nodes 20 and 23, respectively. As shown in Fig.~\ref{fig:reward_convergence_345}, all training curves converge well, indicating that the proposed method maintains stable learning behavior as the number of AIDCs increases. In addition, the achieved reward level improves as more AIDCs are installed in the power distribution system. This implies that the proposed framework can effectively utilize additional resources while maintaining coordinated operation.

Fig.~\ref{fig:reward_345} compares the relative reduction of electricity purchase cost and carbon emissions in the proposed MAT method using the joint mode with respect to the power mode with different numbers of AIDCs. As shown in this figure, the joint mode consistently outperforms the power mode from both economic and environmental perspectives as the number of AIDCs increases.
In addition, the performance gap between the joint and power modes becomes more pronounced  with increasing AIDCs, indicating that the benefit of coordinated operation between the DSO and AIDC operators grows with system size. This result demonstrates that the proposed hierarchical CA-MARL framework not only maintains its effectiveness but also exhibits favorable scalability in coordination efficiency.

Table~\ref{tab:action_complexity} compares the cardinality of action spaces between the non-hierarchical (flat) structure~\eqref{eq:flat_complexity} and the proposed hierarchical structure~\eqref{eq:hier_complexity} as the number of AIDCs increases from $N=2$ to 5. The results in Table ~\ref{tab:action_complexity} are obtained using a discretization resolution $\Gamma=10$, the number of training and inference workload classes $M_{\mathrm{tr}}=2$ and $M_{\mathrm{inf}}=2$, the number of allowable deferral steps $H=6$ for training workloads, and $|\mathcal A^T| = 6$ corresponding to a CRAC supply temperature resolution of $1^\circ$C. As shown in this table, the action-space cardinality of the flat structure grows exponentially with respect to the number of AIDCs, whereas that of the hierarchical structure increases much more gradually. Consequently, the reduction ratio between the flat and hierarchical structures becomes increasingly significant as the system size grows. These results demonstrate that the proposed CA-MARL framework effectively mitigates combinatorial action-space explosion by transforming multiplicative growth into a factorized structure, thereby ensuring scalability and stable learning in multi-AIDC systems.

\begin{table}[t!]
\centering
\caption{Comparison of action-space cardinality between the non-hierarchical (flat) and proposed hierarchical structures with varying number of AIDCs.}
\label{tab:action_complexity}
\renewcommand{\arraystretch}{1.15}
\resizebox{\columnwidth}{!}{%
\begin{tabular}{c c c c}
\noalign{\hrule height 1.2pt}
\textbf{Number of} & \textbf{Non-hierarchical (flat)} & \textbf{Hierarchical} & \textbf{Reduction} \\
\textbf{AIDCs ($N$)} & \textbf{structure} & \textbf{structure} & \textbf{ratio} \\
\noalign{\hrule height 0.8pt}
2 & $4.29\times10^{25}$ & $1.08\times10^{11}$ & $3.96\times10^{14}$ \\
3 & $3.01\times10^{39}$ & $1.62\times10^{11}$ & $1.85\times10^{28}$ \\
4 & $5.75\times10^{52}$ & $2.23\times10^{11}$ & $2.57\times10^{41}$ \\
5 & $4.67\times10^{65}$ & $1.27\times10^{12}$ & $3.67\times10^{53}$ \\
\noalign{\hrule height 1.2pt}
\end{tabular}%
}
\end{table}

\section{Conclusion} \label{sec:conclusion}
In this study, we propose a hierarchical CA-MARL coordination framework
for DSO and AIDC operators to achieve efficient and low-carbon AIDC operations under uncertainties in power distribution systems. Unlike existing studies that utilize location-based signals, the proposed framework employs CEF-induced NCI signals to perform carbon-aware closed-loop DSO--AIDC coordination using the spatial-temporal flexibility of heterogeneous AI training/inference workloads. To this end, the two key features include i) an environment that provides NCI signals using a CEF-integrated DSO problem and ii) a hierarchical MAT-based CA-MARL architecture that prevents combinatorial action-space explosion of global WM and local AIDC agents. The WM and AIDC agents cooperate to ensure efficient and eco-friendly AIDC operation through spatial workload allocation, spatial GPU allocation, and temporal workload shift, along with control of the supply air temperature of the CRAC-based cooling system. To further reduce the dimensionality of the action space of the WM and AIDC agents, the largest remainder rule was adopted to convert the continuous actions associated with the spatial training/inference workload allocation and temporal training workload shift into executable discrete actions.
Numerical examples simulated using an IEEE 33-node system confirmed the effectiveness of the proposed framework using the MAT method in terms of total carbon emissions, carbon emission efficiency, and total training/inference workload throughputs along with the number of dropped training/inference workloads compared to the MAPPO, Transformer, and MAT-Dec methods. 
\bibliographystyle{IEEEtran}
\bibliography{references}

\end{document}